\documentclass[a4paper,fleqn]{cas-sc}
\usepackage[numbers,,sort&compress]{natbib}
\usepackage{caption}
\begin{document}
\let\WriteBookmarks\relax
\def\floatpagepagefraction{1}
\def\textpagefraction{.001}
\shorttitle{Excitable mechanical system under periodic forcing}
\shortauthors{K. Prabith et~al.}

\title [mode = title]{Excitable mechanical lattice: from single-unit bifurcations to boundary-driven pulse propagation}                      

\author[1]{K. Prabith}[orcid=0000-0002-2175-1562]
\ead{prabhithprabha@gmail.com}
\cormark[1]

\credit{Conceptualization, Methodology, Software, Formal Analysis,  Investigation, Visualization, Writing – Original Draft}

\affiliation[1]{organization={Laboratoire d’Acoustique de l’Université du Mans (LAUM)},
addressline={UMR 6613, Institut d’Acoustique - Graduate School (IA-GS), CNRS}, 
city={Le Mans},
postcode={72085}, 
country={France}}

\author[1]{Georgios Theocharis}[orcid=0000-0003-2984-4197]
\credit{Conceptualization, Validation, Supervision, Writing – Review \& Editing}

\author[2]{Sylvain Barbay}[orcid=0000-0003-0092-7829]
\credit{Supervision, Writing – Review \& Editing}
\affiliation[2]{organization={Université Paris-Saclay, CNRS, Centre de Nanosciences et de Nanotechnologies},
addressline={10 Bd Thomas Gobert}, 
city={Palaiseau},
postcode={91120}, 
country={France}}

\author[1]{Soizic Terrien}[orcid=0000-0003-3044-1846]
\credit{Conceptualization, Validation, Supervision, Writing – Review \& Editing, Funding acquisition}

\cortext[cor1]{Corresponding author}

\begin{abstract}
We comprehensively characterize the dynamical regimes of a periodically forced excitable mechanical unit and subsequently investigate how the complex oscillatory responses generated in the first unit under periodic boundary forcing affect pulse propagation through a one-dimensional lattice of such coupled excitable units. Our mechanical unit comprises a mass connected to a fixed frame by an elastic beam and interacting with a moving plate through a smooth, nonmonotonic friction, exhibiting an all-or-none response characteristic of excitable systems. Using two-parameter bifurcation analysis, we identify distinct dynamical regimes in the forcing parameter space and delineate regions of subthreshold and suprathreshold responses associated with intrinsic excitability. Small changes in forcing amplitude can induce rapid transitions between these regimes, with the threshold amplitude depending on the forcing frequency. Within the suprathreshold regime, the system exhibits mixed-mode oscillations (MMOs), characterized by alternating sequences of small-amplitude oscillations (SAOs) and large-amplitude spikes. We then investigate pulse propagation through a lattice of coupled excitable units under sinusoidal boundary forcing and find that the lattice selectively suppresses SAOs and closely spaced large-amplitude spikes within the MMO patterns generated at the first unit. Consequently, the complex boundary dynamics are transformed into pulse trains with nearly uniform spike amplitudes and interspike intervals. When the initial interspike intervals are nonuniform, propagation progressively reorganizes the temporal spacing of transmitted pulses, promoting more uniform intervals in sufficiently long lattices. These findings may also be relevant to other excitable media, where pulse-train regularization during propagation could improve signal transmission even when initial interspike intervals are nonuniform.
\end{abstract}


\maketitle

\section{Introduction}
Excitable systems~\cite{izhikevich2007dynamical} constitute a class of nonlinear dynamical systems characterized by threshold-dependent responses, wherein an external perturbation exceeding a critical threshold drives the system through a large-amplitude excursion in phase space before it relaxes back to its stable resting state, whereas a perturbation below the threshold results in rapid relaxation toward the resting state. Consequently, their response is inherently all-or-none: subthreshold perturbations decay without significant activation, whereas suprathreshold perturbations generate a well-defined pulse that is largely insensitive to the perturbation amplitude. Such dynamics were originally identified in biological processes, including neuronal firing and cardiac excitation~\cite{fitzhugh1961impulses,izhikevich2000neural,barrio2020excitable}, where excitability enables the generation of electrical impulses in response to sufficiently strong stimuli. Analogous excitable behavior has subsequently been demonstrated in a broad range of physical systems, including electronic~\cite{medeiros2012excitable,indiveri2011neuromorphic,wijekoon2008compact}, optical~\cite{prucnal2016recent,dubbeldam1999excitability,goulding2007excitability,barbay2011excitability,garbin2017refractory}, and mechanical~\cite{cartwright1999dynamics} platforms. When coupled to form spatially extended lattices, excitable systems exhibit emergent spatiotemporal dynamics arising from the interplay between local threshold behavior and inter-unit coupling, enabling the propagation of excitation waves through the medium in the form of traveling pulses~\cite{cattani2014fitzhugh,morales2018solitary,alfaro2020pulse}. In biological systems, such pulse propagation constitutes a fundamental mechanism for information transmission and has inspired engineered realizations for energy-efficient signal transport and computation~\cite{adamatzky1998universal,gorecki2009information,pammi2019photonic,soun2022computing}.

Achieving enhanced functionalities in engineered excitable media, however, requires a detailed understanding of how pulse propagation is influenced by the temporal characteristics of external excitation. For instance, subthreshold stimuli that are individually incapable of eliciting an excitable response can collectively reach or exceed the critical threshold through temporal summation when applied to an excitable system at sufficiently short time intervals~\cite{selmi2015temporal}. Moreover, periodic application of suprathreshold stimuli at sufficiently short time intervals can drive the excitable system into forced oscillatory dynamics exhibiting a variety of nonlinear regimes, including frequency locking between intrinsic and external timescales, subharmonic entrainment characterized by rational ratios between the response and forcing frequencies, and bifurcation sequences that may ultimately lead to complex aperiodic dynamics~\cite{chialvo1987non,feingold1988phase,sato1992response,nomura1993bonhoeffer,larotonda2002experimental}. Similar dynamical features have also been observed under sustained periodic forcing~\cite{rajasekar1988period,rajasekarperiod,mendez2001dynamics}, in which the excitable system is continuously driven by a time-periodic input. These rich nonlinear responses arising under periodic forcing motivate a comprehensive characterization of the resulting dynamics across the forcing amplitude--frequency parameter space to identify the distinct dynamical regimes and the boundaries separating them.

An initial attempt in this direction was reported in a recent study~\cite{hellen2025bifurcation}, in which the periodically forced FitzHugh--Nagumo (FHN) model, describing the dynamics of a biological neuron, was analyzed using a two-parameter bifurcation analysis with respect to forcing amplitude and frequency, revealing a rich organization of distinct dynamical regimes. In particular, the study identified isolated, island-like regions of higher-period and aperiodic responses embedded within a background of 1$\times$-periodic oscillations. However, the analysis primarily classified the responses according to their periodicity, without explicitly distinguishing between subthreshold and suprathreshold oscillatory regimes across the forcing parameter space. This distinction is important because intrinsically excitable systems can undergo sharp transitions between these regimes in response to small changes in forcing amplitude. Consequently, a systematic characterization of the dynamical regimes of a periodically forced excitable unit, accounting for both periodicity and oscillation amplitude, is needed to understand how its response can be tuned through external forcing. This tunability may be relevant to engineered excitable lattices driven through a boundary unit, where the amplitude of the boundary unit's oscillations can determine whether excitation is initiated in neighboring units, while the temporal characteristics of its response can influence subsequent pulse propagation through the lattice. These considerations motivate two related questions that remain insufficiently explored: how forcing amplitude and frequency jointly organize the dynamical regimes of an individual excitable unit, including their threshold-dependent features, and how the resulting boundary dynamics influences subsequent pulse propagation in a lattice of coupled excitable units.

\begin{figure}
	\centering
	\includegraphics[width=1\linewidth]{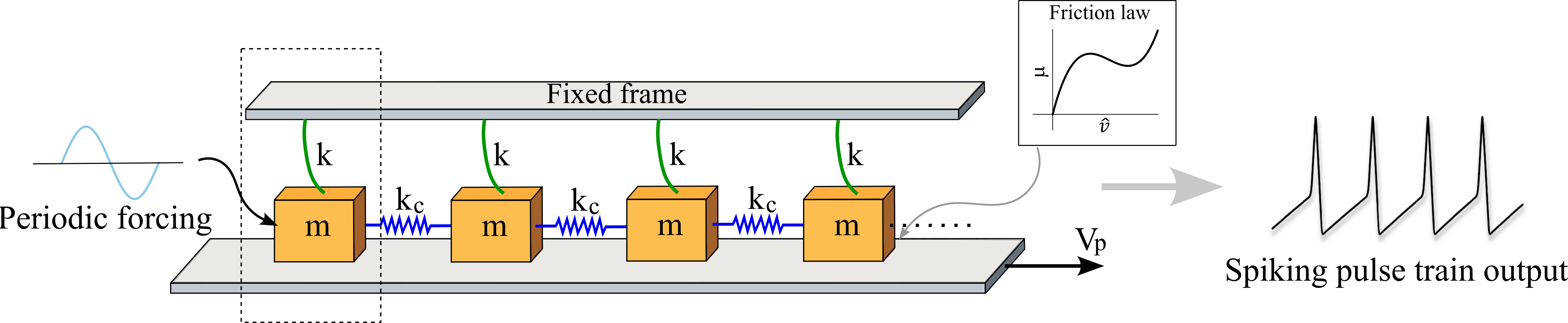}
	\caption{Schematic of the one-dimensional excitable mechanical lattice, consisting of elastically coupled masses connected to a fixed frame by elastic beams and interacting with a moving plate through a nonlinear frictional force. The dashed rectangle highlights a single excitable unit, while the inset shows the smooth, nonmonotonic $N$-shaped friction law responsible for the excitable dynamics. Sinusoidal forcing is applied to the boundary unit, generating pulse trains that propagate through the lattice.}
	\label{Fig1}
\end{figure}

To address these open questions, we first investigate the dynamics of an excitable mechanical unit, governed by creep-slip behavior~\cite{cartwright1997burridge}, subjected to sinusoidal forcing. The unit is inspired by the FHN model~\cite{cebrian2024six} and provides a mechanical framework for investigating periodically driven excitability. It consists of a mass connected to a fixed frame by an elastic beam and interacting with a moving plate through a nonlinear frictional force, as shown within the dashed rectangle in Fig.~\ref{Fig1}. With a smooth, nonmonotonic $N$-shaped friction law, the system exhibits creep-slip behavior, which contributes to its excitable dynamics. The physical realizability of such a friction law is therefore an important consideration. Previous experimental observations suggest that nonmonotonic friction occurs in various materials, including hydrogels confined between contacting surfaces~\cite{kurokawa2005elastic,gong2006friction,shoaib2018insight,shoaib2020advances}. Furthermore, engineered interfaces offer promising routes to tailoring frictional responses. Recent advances in inverse-designed metainterfaces have demonstrated that friction laws can be tailored through the controlled arrangement of surface asperities~\cite{mouton2026friction,aymard2024designing,hartung2023experimental,bruzzone2017engineered}. Magnetic metamaterials provide another possible route to nonmonotonic frictional responses, with collective magnetic reorientations and associated hysteresis potentially producing analogous behavior even in the absence of physical contact~\cite{gu2026non}. These developments motivate the consideration of nonmonotonic friction laws as physically plausible ingredients of excitable mechanical systems and provide a basis for investigating their dynamical consequences under periodic forcing.

We then investigate the spatiotemporal dynamics of a one-dimensional excitable mechanical lattice under sinusoidal forcing applied to its first (boundary) unit. The lattice consists of multiple excitable units coupled elastically, as illustrated in Fig.~\ref{Fig1}. It is mechanically equivalent, although kinematically inverted, to the classical Burridge--Knopoff (BK) model, which was originally introduced to investigate earthquake dynamics~\cite{Burridge1967,carlson1989properties,carlson1991intrinsic,mori2006simulation}. When equipped with a smooth, nonmonotonic $N$-shaped friction law, the BK model provides a prototypical excitable mechanical lattice capable of supporting nonlinear pulse propagation~\cite{morales2016localized}. A practical limitation of the conventional BK configuration is that it is typically realized by dragging the lattice over a stationary plate, which can make controlled laboratory implementation challenging. In contrast, the present configuration enables controlled relative motion between the plate and the masses while maintaining a stationary lattice framework. With this model, we examine how the temporally complex responses generated at the driven boundary unit influence subsequent pulse propagation through the lattice. In particular, we demonstrate that the excitable lattice acts as a nonlinear filter and can asymptotically regularize the pulse propagation dynamics in most cases. Our findings reveal novel wave-propagation mechanism in discrete mechanical lattices and may provide a basis for controlling signal transmission through external forcing.

Our numerical investigation provides grounds for the experimental realization and technological exploitation of periodically driven excitable mechanical systems. More broadly, the physical principles identified here may extend beyond mechanical systems and offer insight into periodically driven excitable dynamics in other classes of excitable media, including electronic and optical platforms. The remainder of this manuscript is organized as follows. In Sec.~\ref{Sec2}, we characterize the response of a single excitable mechanical unit across the forcing amplitude--frequency parameter space. In Sec.~\ref{Sec3}, we extend the analysis to a one-dimensional excitable mechanical lattice driven periodically through its boundary unit and examine how the boundary dynamics influence subsequent pulse propagation through the lattice. Finally, Sec.~\ref{Sec4} summarizes the main findings and discusses directions for future research.

\section{\label{Sec2} Excitable mechanical unit and its dynamics under periodic forcing\protect\\}
We begin by investigating the dynamics of a single excitable mechanical unit subjected to sinusoidal forcing, with the aim of characterizing its response across the forcing amplitude--frequency parameter space and identifying the distinct dynamical regimes that emerge as the forcing parameters vary. The unit consists of a mass $m$ connected to a fixed frame by a slender elastic beam of stiffness $k$ and in frictional contact with a plate moving at a constant velocity $V_p$. Applying Newton's second law, the equation of motion is given by
\begin{equation}
	m \ddot{x} + k x + \mathcal{F}_{fr}(\dot{x} - V_p) = f_{e} \sin(\omega t)
	\label{Eq1}
\end{equation}
where $x(t)$ denotes the displacement of the mass, and overdots indicate derivatives with respect to time $t$. The parameters $f_e$ and $\omega$ denote the amplitude and angular frequency of the external forcing, respectively. The term $\mathcal{F}_{fr}$ represents the nonlinear friction force arising from the relative motion between the mass and the moving plate, expressed as
\begin{equation}
	\mathcal{F}_{fr}(\dot{x} - V_p) = 
	f_0 \, \mu\left(\frac{\dot{x} - V_p}{v_0}\right)\, \mathrm{sgn}\left(\frac{\dot{x} - V_p}{v_0}\right)
	\label{Eq2}
\end{equation}
Here, the function $\mu$ describes the velocity-dependent friction coefficient, while $\mathrm{sgn}$ denotes the sign function, which determines the direction of the friction force relative to the motion of the plate. The parameters $f_0$ and $v_0$ denote characteristic force and velocity scales, respectively. The velocity scale $v_0$ normalizes the relative velocity entering the friction law, whereas $f_0$ sets the overall magnitude of the friction force. In this work, the friction coefficient $\mu$ is assumed to exhibit a smooth, nonmonotonic $N$-shaped dependence on the normalized relative velocity \cite{estrin1996model}. This form introduces a velocity-weakening region between velocity-strengthening regimes, which is essential for the emergence of excitability. Specifically, $\mu$ is modeled by the cubic polynomial,
\begin{equation}
	\mu\left(\hat{v}\right) = a \left|\hat{v}\right|^3 - b \left|\hat{v}\right|^2 + c \left|\hat{v}\right|
	\label{Eq3}
\end{equation}
where $\hat{v}$ denotes a normalized relative velocity. The coefficients $a = 4.8$, $b = 10.1$, and $c = 5.8$ are chosen such that Eq.~\ref{Eq3} yields a smooth $N$-shaped profile consistent with the prescribed frictional behavior. Further details regarding the selection of these coefficients are provided in Appendix~\ref{appA}.

To facilitate the analysis, we introduce the dimensionless displacement $u(t)=x(t)/x_0$, where $x_0=f_0/k$, and the dimensionless time $\tau=t v_0/x_0$. Introducing the dimensionless parameters $\gamma = m v_0^2 k / f_0^2$, $v_p = V_p / v_0$, $F = f_e / f_0$, and $\Omega = \omega x_0 / v_0$, Eq.~\ref{Eq1} can be recast in first-order form as
\begin{eqnarray}
	\frac{du}{d\tau} &=& v, \nonumber \\[2pt]
	\gamma \frac{dv}{d\tau} &=& -u - \Phi_{fr}(v - v_p) + F \sin(\Omega \tau)
	\label{Eq4}
\end{eqnarray}
where $\Phi_{fr}(v - v_p) = \mu(v - v_p)\mathrm{sgn}(v - v_p)$. In this form, the system constitutes a multiple-timescale dynamical system with a clear fast–slow structure for $\gamma \ll 1$, with $u$ evolving on the slow timescale and $v$ on the fast timescale. In the present system, this condition can be achieved through appropriate parameter selection, in particular by employing very light masses, very soft springs, and/or sufficiently large values of $f_0$.

\subsection{Excitability characteristics}
\begin{figure}
	\centering
	\includegraphics[width=0.75\linewidth]{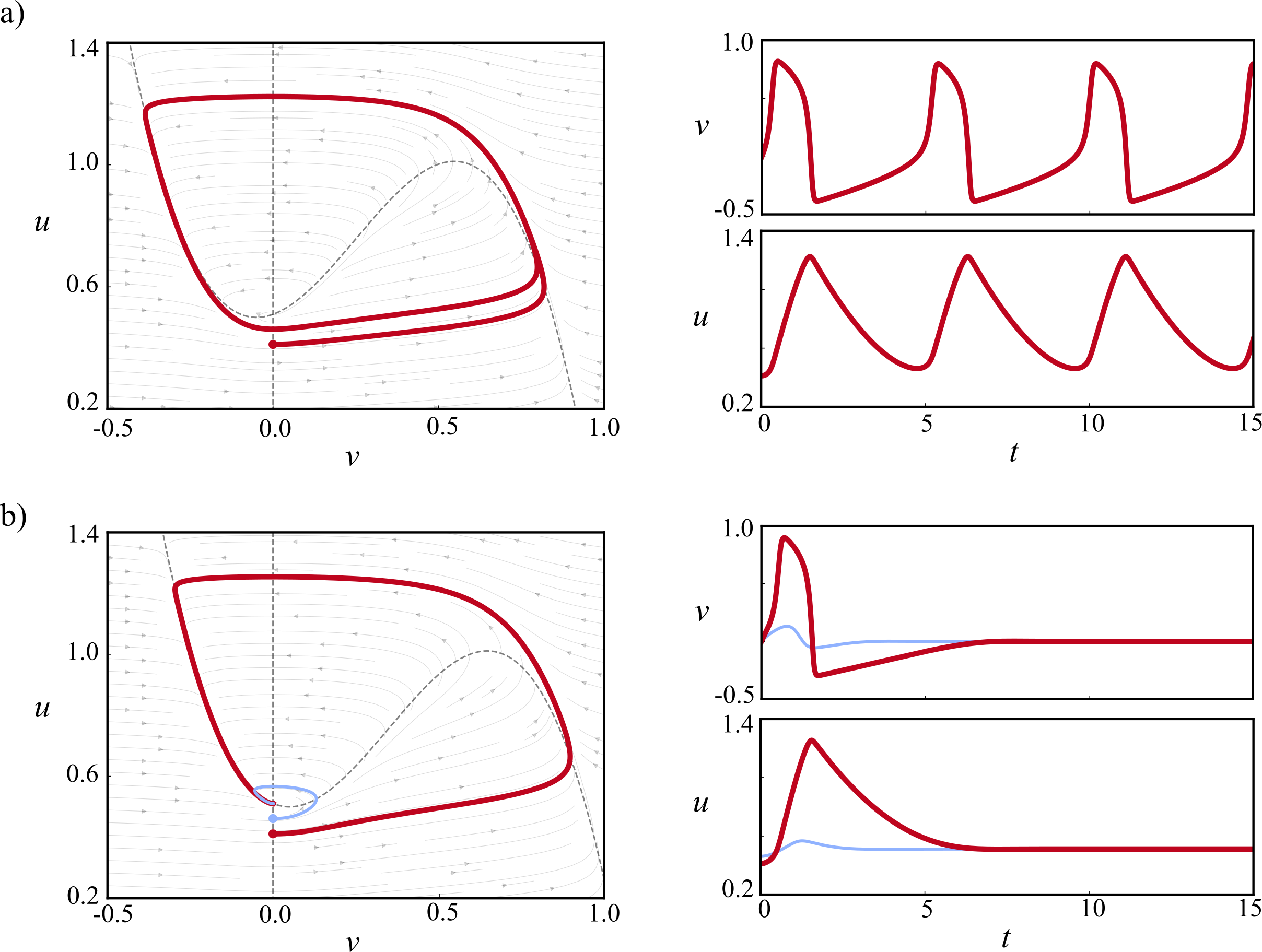}
	\caption{\label{Fig2} Phase-space trajectories and corresponding time evolution of the system variables for $\gamma=0.15$ and two values of $v_p$: (a) $v_p=0.95$ and (b) $v_p=1.05$. Panel (a) shows relaxation oscillations on a stable limit cycle, whereas panel (b) illustrates excitability. In panel (b), a subthreshold perturbation (light blue) decays rapidly to the stable fixed point, whereas a suprathreshold perturbation (red) produces a large excursion and a sharp transient pulse before returning to equilibrium.}
\end{figure}

To reveal the intrinsic excitability of the mechanical unit, we first examine its dynamics in the absence of external forcing. This provides a baseline for understanding the system's response to perturbations. Setting the forcing amplitude to zero in Eq.~\ref{Eq4} yields the autonomous dynamical system
\begin{eqnarray}
	\frac{du}{d\tau} &=& v, \nonumber \\[2pt]
	\gamma \frac{dv}{d\tau} &=& -u - \Phi_{fr}(v - v_p)
	\label{Eq5}
\end{eqnarray}
The dynamical structure of this system can be examined using its nullclines, which are defined by the conditions $du/d\tau=0$ and $dv/d\tau=0$. These are given by
\begin{equation}
	\begin{aligned}
		v &= 0, && \hspace{1cm} \text{$u$-nullcline},\\
		u &= -\Phi_{fr}(v-v_p), &&  \hspace{0.96cm} \text{$v$-nullcline}.
	\end{aligned}
	\label{Eq6}
\end{equation}
The $u$-nullcline is a vertical line in the $(v,u)$ plane, whereas the shape of the $v$-nullcline is determined by the nonlinear friction law. For the cubic friction law considered here, the $v$-nullcline has an S-shaped profile with two outer branches separated by a middle branch, distinguished by their slopes [see Fig.~\ref{Fig2}]. The intersections of the two nullclines determine the fixed points of the system. Since the $u$-nullcline is defined by $v=0$, the fixed point is given by $(u^*,v^*)=(-\Phi_{fr}(-v_p),0)$. Physically, this fixed point corresponds to a steady-slip state in which the mass remains stationary while the plate slides beneath it, with the spring force balancing the frictional force. The linear stability analysis of this fixed point yields the eigenvalues
\begin{equation}
	\lambda_{\pm}=
	\frac{-\Phi_{fr}'(-v_p)\pm
		\sqrt{\left[\Phi_{fr}'(-v_p)\right]^2-4\gamma}}
	{2\gamma}
	\label{Eq9}
\end{equation}
where $\Phi_{fr}'(-v_p)$ enotes the slope of the friction characteristic evaluated at the equilibrium. When the fixed point lies on the middle branch of the S-shaped fast nullcline, $\Phi_{fr}'(-v_p)<0$, the real parts of both eigenvalues are positive, and the fixed point is locally unstable. In contrast, when the fixed point lies on either of the outer branches, $\Phi_{fr}'(-v_p)>0$, both eigenvalues have negative real parts, indicating a locally stable equilibrium. For the cubic friction law adopted in this work, $\Phi_{fr}'(-v_p)$ changes sign at $|v_p|=1$. Accordingly, the fixed point lies on the middle, locally unstable branch of the $v$-nullcline for $|v_p|<1$ and on an outer, locally stable branch for $|v_p|>1$. The corresponding dynamical responses are illustrated in Fig.~\ref{Fig2}. For $v_p=0.95$, the fixed point is unstable, and the system approaches a stable limit cycle corresponding to self-sustained relaxation oscillations [Fig.~\ref{Fig2}(a)]. In contrast, for $v_p=1.05$, the fixed point is stable [Fig.~\ref{Fig2}(b)]. A subthreshold perturbation produces a small-amplitude response that decays rapidly toward the fixed point [light-blue trajectory in Fig.~\ref{Fig2}(b)]. When the perturbation amplitude is slightly increased, the trajectory makes a large excursion before returning to the same fixed point [red trajectory in Fig.~\ref{Fig2}(b)], producing a pronounced transient pulse in $v$. Physically, this large excursion corresponds to a rapid displacement of the mass on the moving plate, followed by a return to the steady-slip (equilibrium) state. These responses highlight the excitable nature of the mechanical unit, which is analogous to the FHN model, as discussed in Appendix~\ref{appB}.

Throughout the manuscript, we fix the parameters at $\gamma=0.15$ and $v_p=1.05$. The value $\gamma=0.15$ is chosen to provide a clear separation between the fast and slow timescales while remaining within a physically relevant parameter range for the mechanical system considered. The choice $v_p=1.05$ places the fixed point just beyond the critical value $|v_p|=1$, with the stable equilibrium located close to the fold of the fast nullcline. This proximity ensures a well-defined excitable regime in which relatively small perturbations can trigger large-amplitude excursions.

\subsection{Dynamical regimes in the forcing parameter space}
We now examine the dynamical regimes of the periodically forced excitable mechanical system ($F\neq0$), focusing on how the forcing amplitude $F$ and frequency $\Omega$ govern its long-time behavior. Under external forcing, the excitable system becomes a nonautonomous, periodically forced oscillator. Unlike a conventional periodically forced oscillator, the present system exhibits excitable characteristics: an increase in the forcing amplitude can induce a qualitative transition from a subthreshold oscillatory response to a suprathreshold oscillatory response involving an excitable excursion. To characterize this transition and identify the boundary separating these two dynamical regimes, we examine the forced dynamics in the $(\Omega,F)$ parameter space.

\begin{figure}
	\centering
	\includegraphics[width=0.62\linewidth]{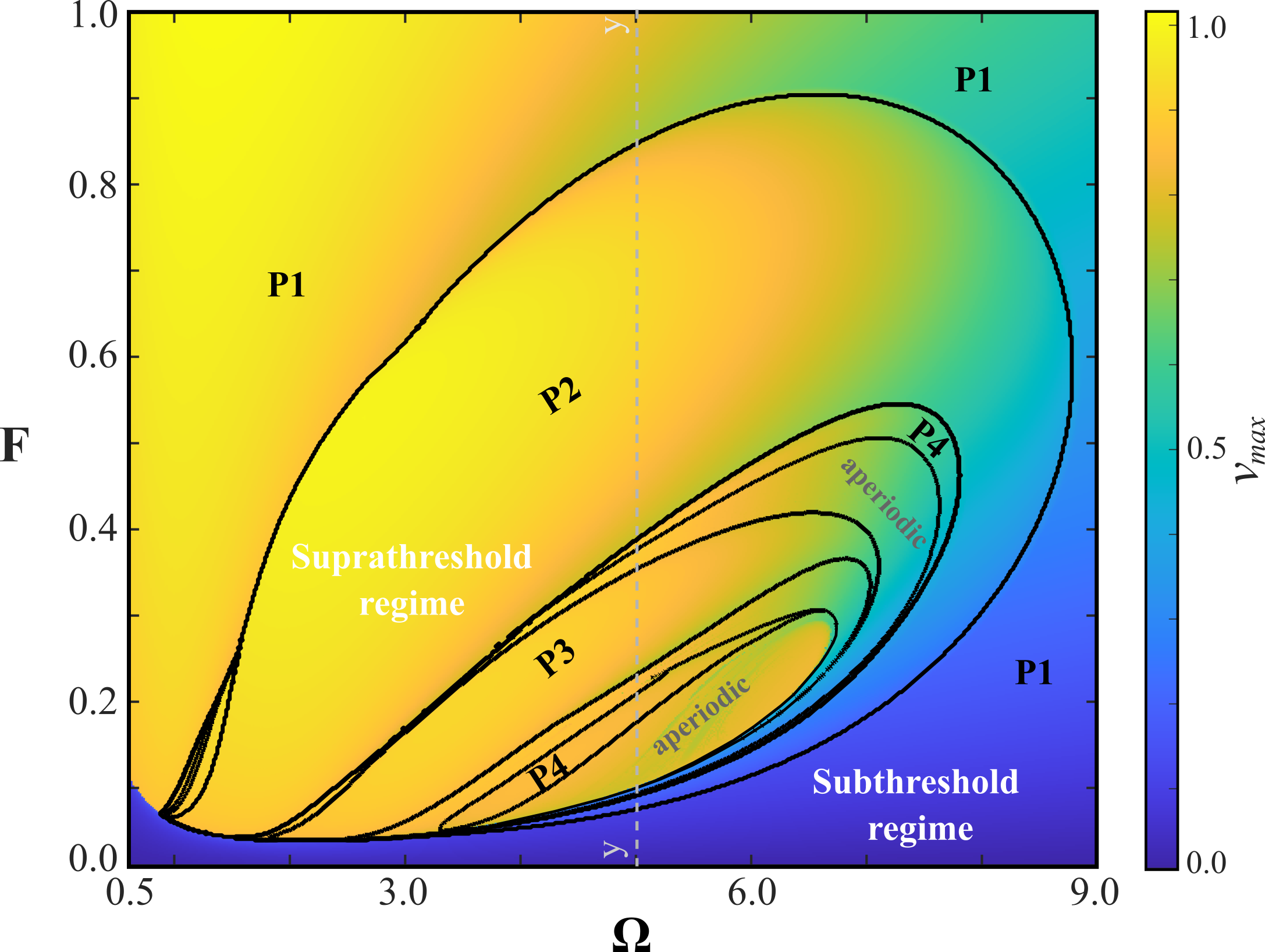}
	\caption{\label{Fig3} Response landscape of the periodically forced excitable mechanical unit in the $(\Omega,F)$ parameter space. The color map shows the maximum velocity $v_{\max}$ of the asymptotic response, highlighting the transition between subthreshold and suprathreshold oscillations as a function of the forcing frequency. Black curves delineate the boundaries between periodic responses, with the regions labeled $P1$--$P4$ corresponding to responses that repeat after one to four forcing periods, respectively. Regions exhibiting aperiodic responses are also indicated. The dashed gray line indicates the path at fixed forcing frequency $\Omega=5$, along which the forcing amplitude $F$ is varied to examine the bifurcations and transitions between dynamical regimes, as discussed in Fig.~\ref{Fig4}.}
\end{figure}

For each point in the $(\Omega,F)$ parameter space, Eqs.~\ref{Eq4} are integrated numerically for a sufficiently long time, with the initial condition taken as the equilibrium state of the unforced system. Following the decay of transient dynamics, the resulting asymptotic response is used to characterize the forced dynamics. The response amplitude is quantified by the maximum value of the fast variable, $v_{\max}$, which provides a direct measure of the large-amplitude excursion characteristic of the excitable response and thereby distinguishes suprathreshold oscillations from subthreshold oscillations. The resulting response landscape is shown in Fig.~\ref{Fig3}, where the color scale represents $v_{\max}$ in the $(\Omega,F)$ parameter space. At low forcing amplitudes, the response remains confined to subthreshold oscillations, as indicated by the low values of $v_{\max}$. With increasing $F$, a rapid transition to suprathreshold oscillations occurs over a broad range of forcing frequencies, particularly for $0.5\lesssim\Omega\lesssim6.6$. The threshold amplitude varies nonmonotonically with the forcing frequency, reflecting the interplay between the intrinsic timescale of the excitable system and the timescale of the external forcing.

At very low frequencies, the forcing period is long compared with the intrinsic timescale of the system, and a relatively large forcing amplitude is required to drive the trajectory away from its stable equilibrium and trigger an excitable excursion. As the frequency increases, the forcing becomes more effective at inducing large-amplitude excursions, resulting in a decrease in the threshold amplitude over $0.5\lesssim\Omega\lesssim2$. The threshold amplitude then remains approximately constant over $2\lesssim\Omega\lesssim3$. For $3\lesssim\Omega\lesssim6$, the threshold amplitude increases progressively with frequency. This increase can be attributed to the shortening of the forcing period relative to the intrinsic timescale of the system. As the forcing period becomes shorter, successive forcing cycles arrive before the system has sufficiently recovered from the preceding excursion. Consequently, a larger forcing amplitude is required to induce the suprathreshold oscillatory response as the frequency increases. At still higher frequencies, the forcing varies so rapidly that the system cannot respond coherently to individual forcing cycles, further reducing its effectiveness in inducing large-amplitude excursions. As a result, the transition to the suprathreshold response becomes less pronounced, and the dynamics become increasingly dominated by the direct effect of the external forcing. This behavior is evident in Fig.~\ref{Fig3} for $\Omega\gtrsim6.6$, where the boundary separating the subthreshold and suprathreshold responses becomes progressively less distinct with increasing frequency.

In addition to distinguishing between subthreshold and suprathreshold responses, Fig.~\ref{Fig3} characterizes the temporal organization of the asymptotic oscillations through a stroboscopic Poincar\'e analysis. The black curves delineate the boundaries between responses with different periods, with the regions labeled $P1$--$P4$ corresponding to responses that repeat after one to four forcing periods, respectively; aperiodic regions are also indicated. The $P1$ state forms the background of the parameter space and corresponds to synchronization with the external forcing, whereas the higher-period states represent subharmonic locking, with the response repeating only after multiple forcing cycles. These higher-period states form island-like domains within the $P1$ background, with the $P2$ state extending along major portions of their boundaries, while higher-order subharmonic states are interspersed with regions of aperiodic dynamics. For clarity, Fig.~\ref{Fig3} displays periodic states only up to $P4$; an extended regime map in Appendix~\ref{appC} reveals additional higher-order periodic states, with responses identified up to $P8$. The aperiodic dynamics may include chaotic responses. In the present study, possible chaotic behavior is assessed only qualitatively, based on the corresponding time series and phase-space trajectories, without quantitatively determining Lyapunov exponents.
\begin{figure}
	\centering
	\includegraphics[width=0.68\linewidth]{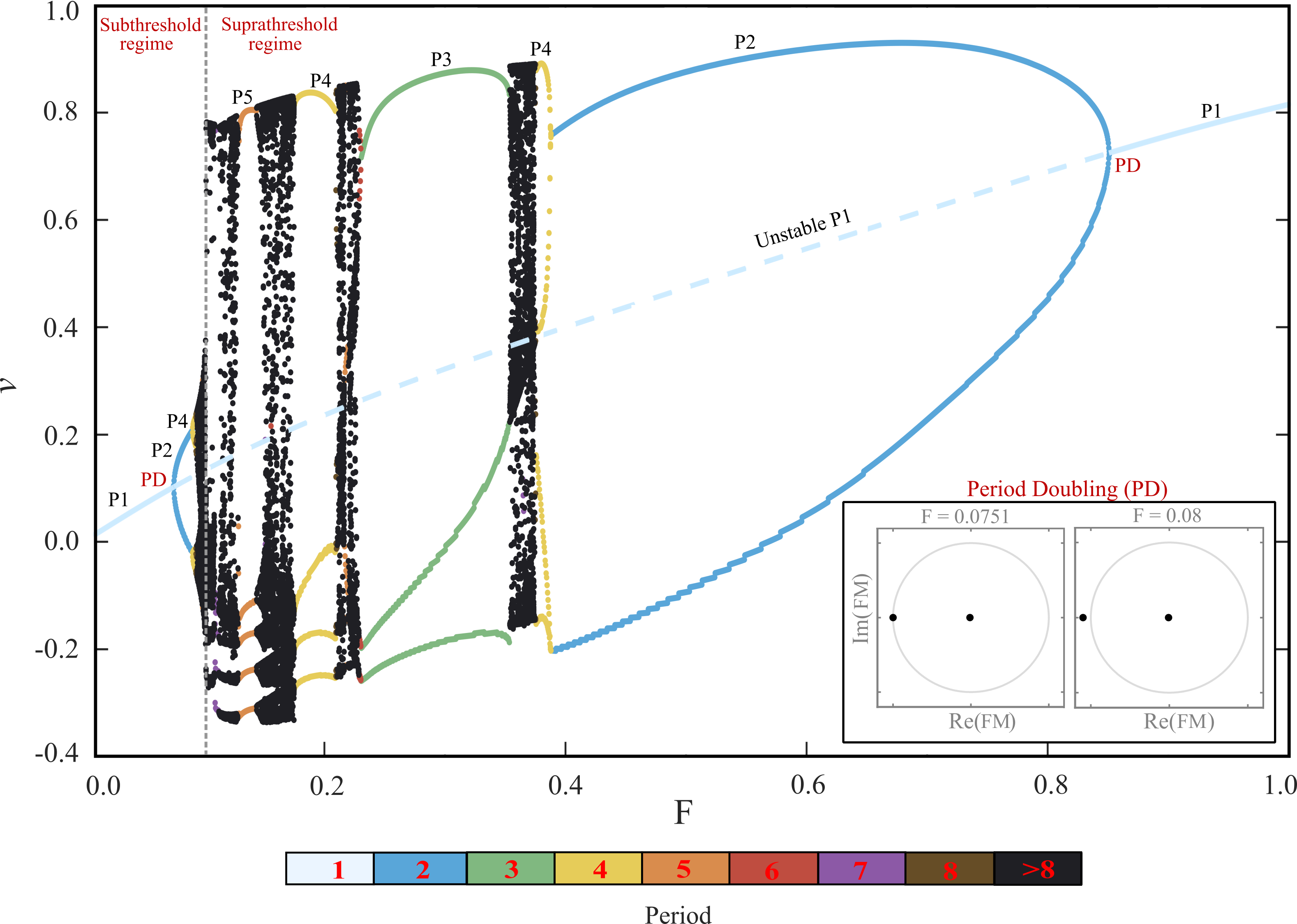}
	\caption{\label{Fig4} Stroboscopic Poincar\'e bifurcation diagram of a single excitable mechanical unit obtained by monotonically increasing the forcing amplitude $F$ at a fixed forcing frequency $\Omega = 5$. The color scale indicates the response period relative to the forcing period. The unstable $P1$ branch obtained via pseudo-arclength continuation is shown by light-blue dashed line. The inset shows the corresponding Floquet multipliers, indicating that the loss of stability of the $P1$ solution occurs through a period-doubling bifurcation, as one multiplier exits the unit circle through the negative real axis.}
\end{figure}

Having established the overall response landscape in the $(\Omega,F)$ parameter space, we next examine in greater detail the bifurcations and transitions between distinct dynamical regimes as the forcing amplitude is varied. To this end, we fix the forcing frequency at $\Omega=5$ and vary the forcing amplitude $F$ along the dashed gray line in Fig.~\ref{Fig3}. The resulting stroboscopic Poincar\'e diagram, shown in Fig.~\ref{Fig4}, is obtained by sampling the fast variable $v$ once every forcing period $T$, following the decay of transients, with the sampling initiated at the maximum of $v$. Different response periods are represented by distinct colors, while periodic responses with $P>8$ and aperiodic responses are indicated in black. This construction provides a detailed characterization of the evolution of the system response and the transitions between dynamical regimes with increasing forcing amplitude.

\begin{figure}
	\centering
	\includegraphics[width=1\linewidth]{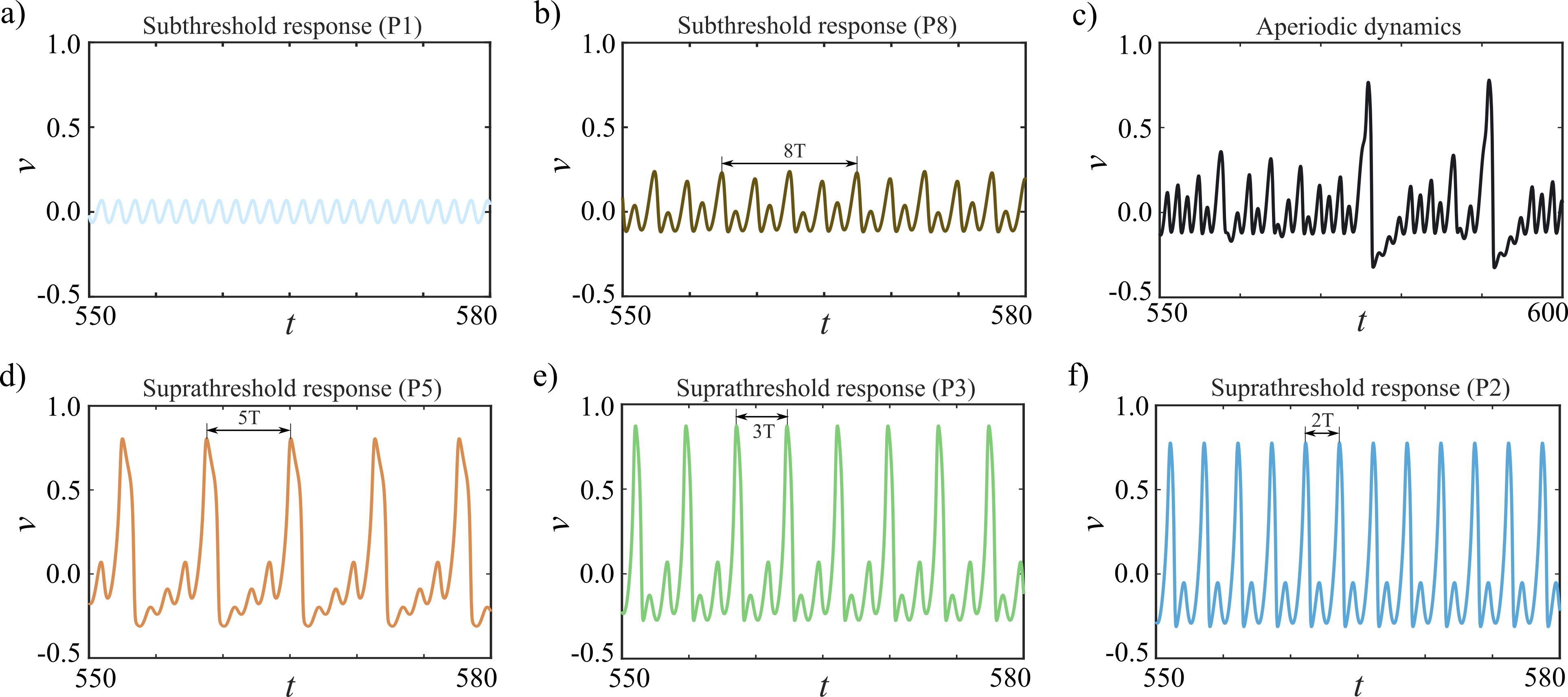}
	\caption{\label{Fig5} Velocity time histories of a single excitable mechanical unit under sinusoidal forcing with fixed forcing frequency $\Omega = 5$. Panels (a)--(f) correspond to forcing amplitudes $F = 0.05$, $0.0938$, $0.1021$, $0.14$, $0.3$, and $0.4$, respectively. Panels (a) and (b) exhibit subthreshold responses, panel (c) shows suprathreshold aperiodic oscillations, and panels (d)--(f) display suprathreshold MMOs characterized by $1^3$, $1^2$, and $1^1$ patterns.}
\end{figure}

For $F \leq 0.0751$, the system exhibits a stable $P1$ response. The corresponding velocity profile at $F = 0.05$ [Fig.~\ref{Fig5}(a)] demonstrates that the oscillations remain subthreshold and do not produce spike-like events. In this regime, the mass undergoes smooth small-amplitude oscillations about the equilibrium state without exceeding the threshold necessary to initiate rapid slip motion. As the forcing amplitude is increased, a period-doubling bifurcation occurs at $F = 0.0751$, leading to the emergence of a stable $P2$ response. This transition is also identified via pseudo-arclength continuation with respect to $F$, as shown by the light-blue curve in Fig.~\ref{Fig4} (solid line: stable; dashed line: unstable). It is further characterized by the corresponding Floquet multipliers (FM), computed along the same branch. The inset of Fig.~\ref{Fig4} shows the FMs at the bifurcation point $F = 0.0751$ and at $F = 0.08$. It is seen that at the bifurcation point, one multiplier exits the unit circle through the negative real axis, confirming a period-doubling instability. Notably, all higher-period and aperiodic responses are found to reside within the parameter region where the $P1$ branch is unstable. 

The stable $P2$ response persists up to $F=0.09$. Upon further increase in $F$, the system undergoes a cascade of successive period-doubling transitions ($P4$, $P8$, \textit{etc.}), ultimately leading to aperiodic oscillations. Representative velocity time series are shown in Figs.~\ref{Fig5}(b) and \ref{Fig5}(c) for $F = 0.0938$ and $F = 0.1021$, corresponding to subthreshold $P8$ and aperiodic responses, respectively. As evident from Fig.~\ref{Fig5}(c), the system exhibits spike-like events, wherein the mass undergoes intermittent rapid slip motions interspersed with intervals of slow evolution near equilibrium. This transition from subthreshold to suprathreshold regime is also indicated in Fig.~\ref{Fig4} by a dashed line. Owing to the aperiodic nature of the response, this behavior may be interpreted as chaotic spiking in excitable mechanical unit, analogous to that observed in periodically driven neuronal systems~\cite{rajasekar1988period,feingold1988phase,rajasekarperiod}. The system remains in the chaotic spiking regime for $0.1021 \leq F \leq 0.128$, with intermittent narrow windows of $P7$ and higher-period ($P > 8$) responses embedded within the aperiodic dynamics. Beyond $F = 0.128$, the system transitions to a $P5$ response that persists up to $F = 0.143$. Figure~\ref{Fig5}(d) shows the corresponding velocity time history at $F = 0.14$, exhibiting a characteristic pattern of one large-amplitude spike followed by three small-amplitude oscillations. Such responses are referred to as mixed-mode oscillations (MMOs)~\cite{desroches2012mixed}, consisting of alternating sequences of large-amplitude spikes and small-amplitude oscillations (SAOs) within a single response period. These patterns are conventionally denoted by $L^s$, where $L$ and $s$ represent the number of large-amplitude spikes and SAOs, respectively. Accordingly, the observed response corresponds to an MMO of the form $1^3$, reflecting the coexistence of distinct amplitude scales within the dynamics. Further examination of the velocity time histories for different forcing amplitudes reveals additional MMOs of the type $1^s$, including $1^2$ and $1^1$, as illustrated in Figs.~\ref{Fig5}(e) and \ref{Fig5}(f), respectively. Importantly, a major portion of the island-like structure in parameter space exhibits MMOs of the form $1^s$.

Beyond $F = 0.143$, the $P5$ response loses stability through a period-doubling bifurcation, giving rise to aperiodic dynamics. This aperiodic regime persists up to $F \approx 0.175$, suggesting potentially chaotic behavior. The system then transitions to a $P4$ response, which remains stable over the interval $0.175 \leq F \leq 0.209$. An MMO of type $1^3$ also exists in this region. As $F$ increases, the dynamics again become aperiodic, interspersed with higher-periodic responses, followed by the emergence of a $P3$ regime over the interval $0.231 \leq F \leq 0.353$. This regime contains an MMO of type $1^2$, as shown in Fig.~\ref{Fig5}(e). The system again enters an aperiodic state at $F \approx 0.353$. Around $F \approx 0.375$, a reverse sequence of bifurcations is observed, wherein the aperiodic dynamics give way to periodic responses through a cascade of reverse period-doubling transitions, successively yielding $P4$, $P2$, and $P1$ oscillations. The velocity time series at $F = 0.4$ exhibits an MMO of type $1^1$ associated with the $P2$ response, as shown in Fig.~\ref{Fig5}(f). This reverse sequence of period-doubling transitions reflects a progressive restoration of synchronous behavior with increasing $F$.

\begin{figure}
	\centering
	\includegraphics[width=0.7\linewidth]{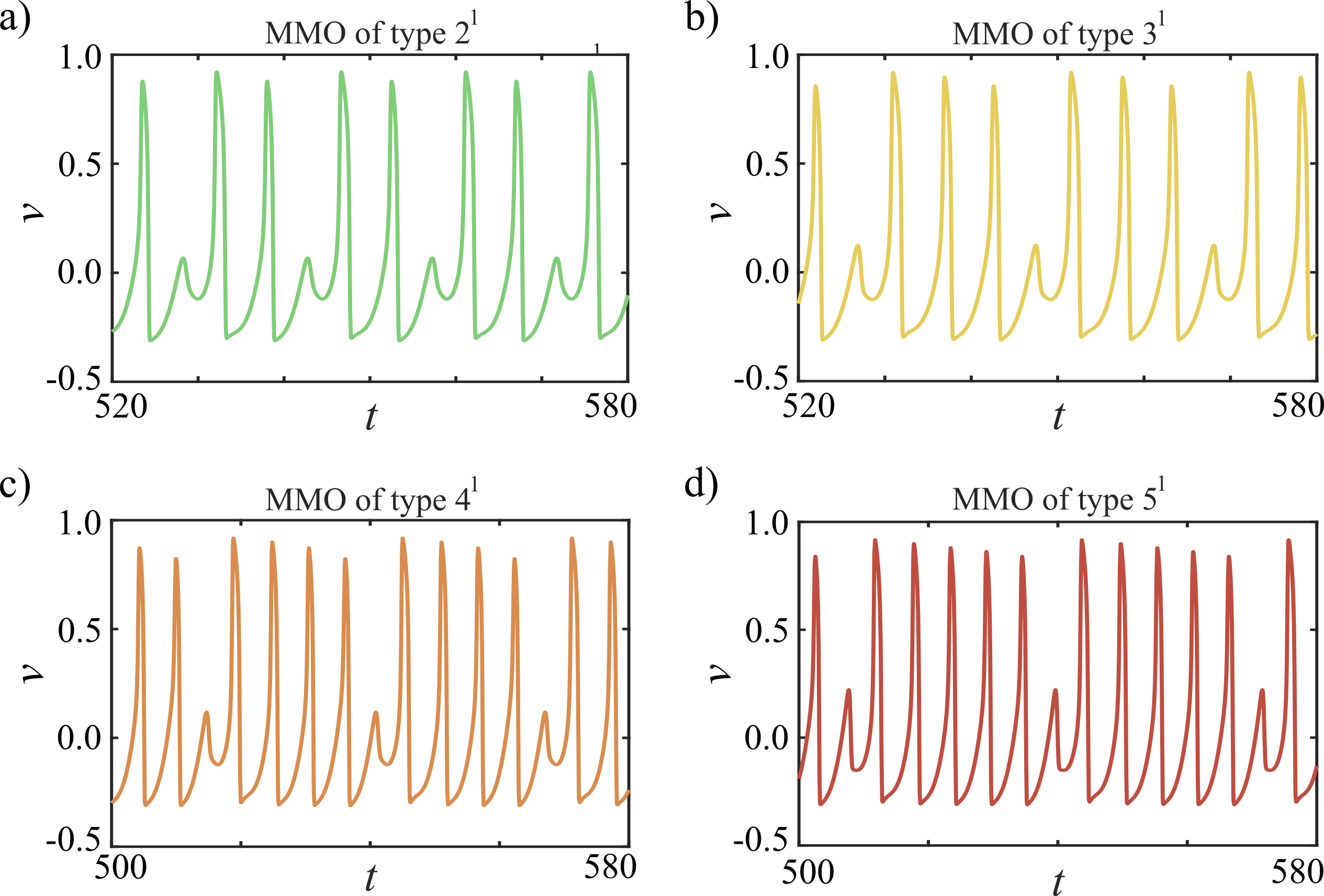}
	\caption{\label{Fig6} Velocity time histories of a single excitable mechanical unit at $F = 0.14$ for different forcing frequencies $\Omega$, taken from the lower-left region of the island-like structure in the response landscape. Panels (a)--(d) correspond to MMO patterns $2^1$, $3^1$, $4^1$, and $5^1$, illustrating a different $L^1$-type sequence compared to the previously observed $1^s$-type MMOs.}
\end{figure}

Another important feature of the system is the emergence of a different pattern of MMOs in the lower-left corner of the island-like structure in Fig.~\ref{Fig3}, which is also shown in Appendix~\ref{appC}. Velocity time series corresponding to representative forcing parameters $(\Omega, F)$ within this region are presented in Fig.~\ref{Fig6}, where the forcing amplitude is fixed at $F = 0.14$ and the forcing frequency $\Omega$ is varied. These time histories reveal MMO patterns of the form $L^1$, specifically $2^1$, $3^1$, $4^1$, and $5^1$, characterized by multiple large-amplitude spikes followed by a single small-amplitude oscillation within a single response period. This behavior differs from the previously discussed $1^s$-type MMOs, demonstrating that the system can support alternative MMO sequences depending on the location within the response landscape. This observation underscores the richness of the dynamical behavior even within small regions of the forcing parameter space.

The occurrence of MMOs in the present system originates from its nature as a nonautonomous, multiple-timescale dynamical system, effectively described by the slow variable $u$, the fast variable $v$, and the phase of the external forcing. The periodic forcing modulates the system's excitable dynamics, giving rise to complex oscillatory patterns that combine small-amplitude oscillations with large-amplitude excitable excursions.  The resulting MMO patterns therefore reflect the combined effects of the system’s intrinsic excitability and the external periodic forcing. The observed dynamics of the sinusoidally driven excitable mechanical unit closely resemble those reported for the periodically forced FHN model~\cite{hellen2025bifurcation}. This similarity highlights the generic role of fast-slow dynamics and periodic modulation in organizing complex oscillatory responses in excitable systems. These observations provide a foundation for understanding how periodic boundary forcing affects pulse propagation in a lattice of coupled excitable units. In particular, we investigate how the temporally structured responses of the boundary unit influence the pulse propagation characteristics in the following section.

\section{\label{Sec3} Pulse propagation in an excitable mechanical  lattice under periodic boundary forcing\protect\\}
We now extend the analysis to a one-dimensional lattice of elastically coupled excitable mechanical units [Fig.~\ref{Fig1}] and investigate the characteristics of pulse propagation under sinusoidal forcing applied to the first unit. The coupling introduces spatially extended dynamics through nearest-neighbor interactions, thereby capturing the interplay between local excitability and inter-unit coupling. The nondimensionalized equations governing the $j$th unit are
\begin{eqnarray}
	\frac{du_j}{d\tau} &=& v_j, \nonumber \\[2pt]
	\gamma \frac{dv_j}{d\tau} &=& -u_j + \beta \, \Delta_d u_j - \Phi_{fr}(v_j - v_p)
	+ \delta_{j1} F \sin(\Omega \tau) \hspace{5mm}
	\label{Eq10}
\end{eqnarray}
where $j = 1,2,\dots,N$, and $\beta = k_c/k$ denotes the dimensionless coupling stiffness. Here, $\Delta_d u_j$ represents the discrete Laplacian associated with nearest-neighbor interactions along the lattice and it is defined as $\Delta_d u_j = u_{j+1} - 2u_j + u_{j-1}$. The term $\delta_{j1}$ denotes the Kronecker delta, ensuring that the external forcing is applied exclusively to the first unit. The lattice is subject to free--free boundary conditions at both ends. The coupling term $\beta\Delta_d u_j$ mediates the transfer of excitation between neighboring units, thereby enabling the propagation of localized pulses through the lattice. Its strength, controlled by the dimensionless coupling parameter $\beta$, determines whether the excitation generated at the boundary unit can successfully activate neighboring units and develop into a propagating pulse. In particular, a critical coupling strength may exist, below which the excitation fails to propagate through the lattice and above which sustained pulse propagation becomes possible \cite{morales2018solitary}. In the present analysis, we consider $\beta = 0.2$ for a chain of $N=20$ coupled excitable units, which is above the critical coupling strength required for successful pulse propagation along the lattice in response to a single perturbation.

Under periodic boundary forcing, the first unit of the lattice exhibits several dynamical features that closely resemble those observed for a single excitable  unit under periodic forcing, while also displaying notable differences induced by coupling. Depending on the forcing parameters, the first unit undergoes subthreshold and suprathreshold oscillations and exhibits MMOs under suprathreshold regime. The corresponding parameter space displays a considerably richer structure, characterized by a greater prevalence of higher-period responses induced by coupling between neighboring units (see Appendix~\ref{appD}). These local dynamics of the first unit play a central role in determining the characteristics of pulse propagation along the lattice. In the following subsections, we examine how pulse propagation is influenced by the distinct dynamical states of the first unit, with particular emphasis on different MMO patterns.

\subsection{$1^s$-type MMOs at the first unit}\label{subsecA}
\begin{figure}
	\centering
	\includegraphics[width=0.55\linewidth]{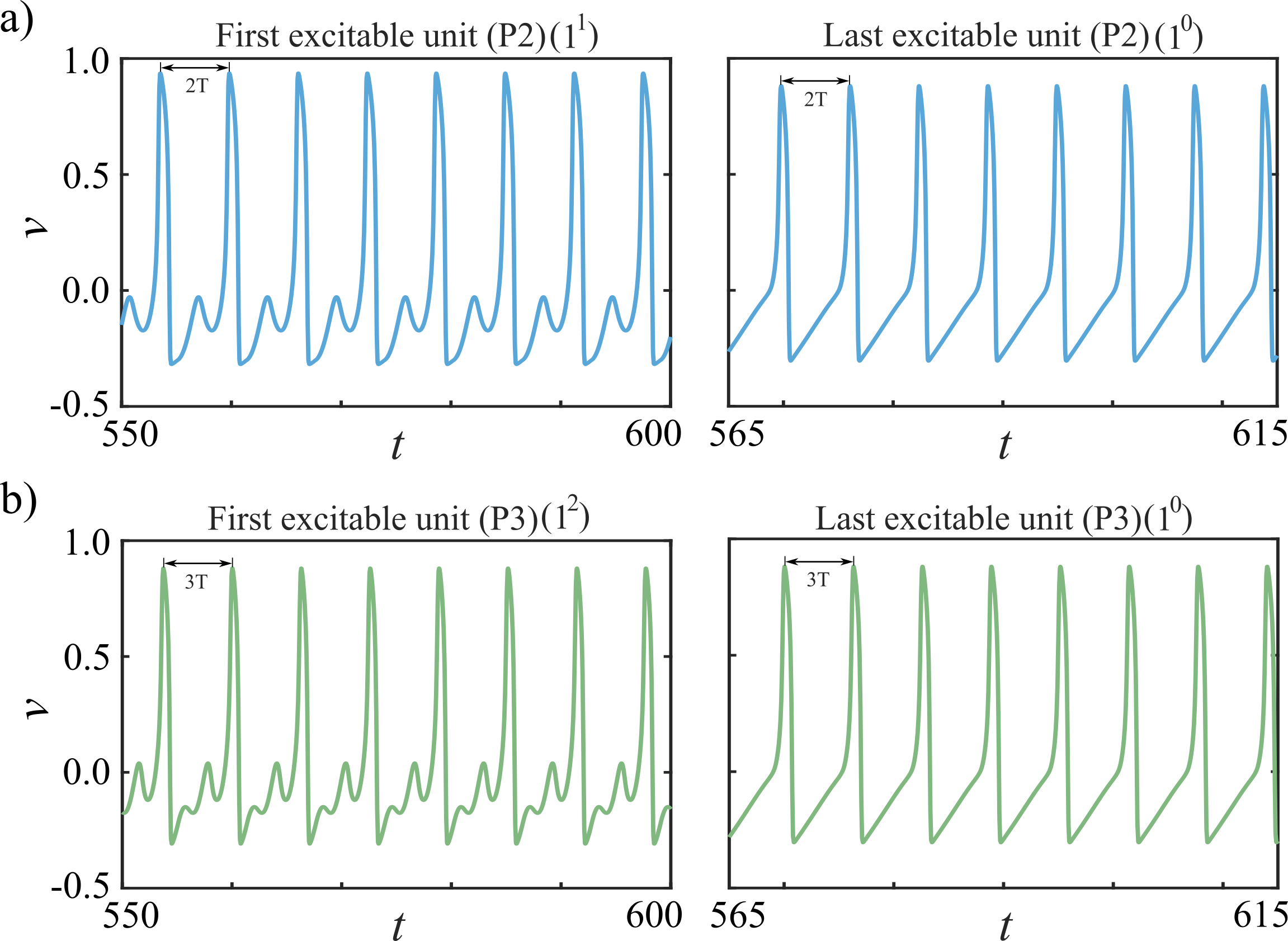}
	\caption{\label{Fig7} Temporal velocity profiles of the first and last units in a one-dimensional lattice of coupled excitable mechanical units ($N = 20$) for $1^{s}$-type MMO states generated at the first unit. Panels (a) and (b) correspond to the forcing parameters $(\Omega,F) = (2,0.2)$ and $(3,0.1)$, representing the $1^1$ and $1^2$ MMO states, respectively. The response at the first unit exhibits MMOs characterized by the coexistence of large-amplitude spikes and SAOs. In contrast, the dynamics at the last unit are dominated exclusively by large-amplitude spikes, while the SAOs are suppressed during propagation.}
\end{figure}

We begin by examining pulse propagation in the lattice when the first excitable unit exhibits $1^{s}$-type MMOs. The resulting sequence of excitation pulses, referred to as a pulse train, propagates through the lattice and carries the temporal structure imposed by the boundary unit. To characterize the resulting spatiotemporal evolution, we compare the dynamical responses of the first and last excitable units, thereby quantifying how the MMO pattern generated at the first unit is modified during transmission through the lattice. Figures~\ref{Fig7}(a) and \ref{Fig7}(b) present the temporal velocity profiles of the first and last excitable units corresponding to the $1^1$ and $1^2$ MMO states, respectively. As shown in Fig.~\ref{Fig7}(a)--(b), the response of the first unit consists of a sequence of uniformly spaced large-amplitude spikes interspersed with SAOs. In contrast, the response at the last unit is significantly simplified, with only the large-amplitude spikes persisting while the SAOs are entirely suppressed. This selective transmission arises from the threshold-dependent activation and refractory dynamics intrinsic to each excitable unit. Following a large-amplitude excitation, an individual unit enters a refractory period, during which its excitability is temporarily reduced, making it less responsive to subsequent stimulation. During this period, the small-amplitude oscillations are insufficient to trigger large-amplitude excitations in downstream units. Consequently, these excursions are progressively suppressed along the lattice, whereas the well-separated large-amplitude spikes exceed the excitation threshold and propagate to the last unit.

\begin{figure}
	\centering
	\includegraphics[width=0.65\linewidth]{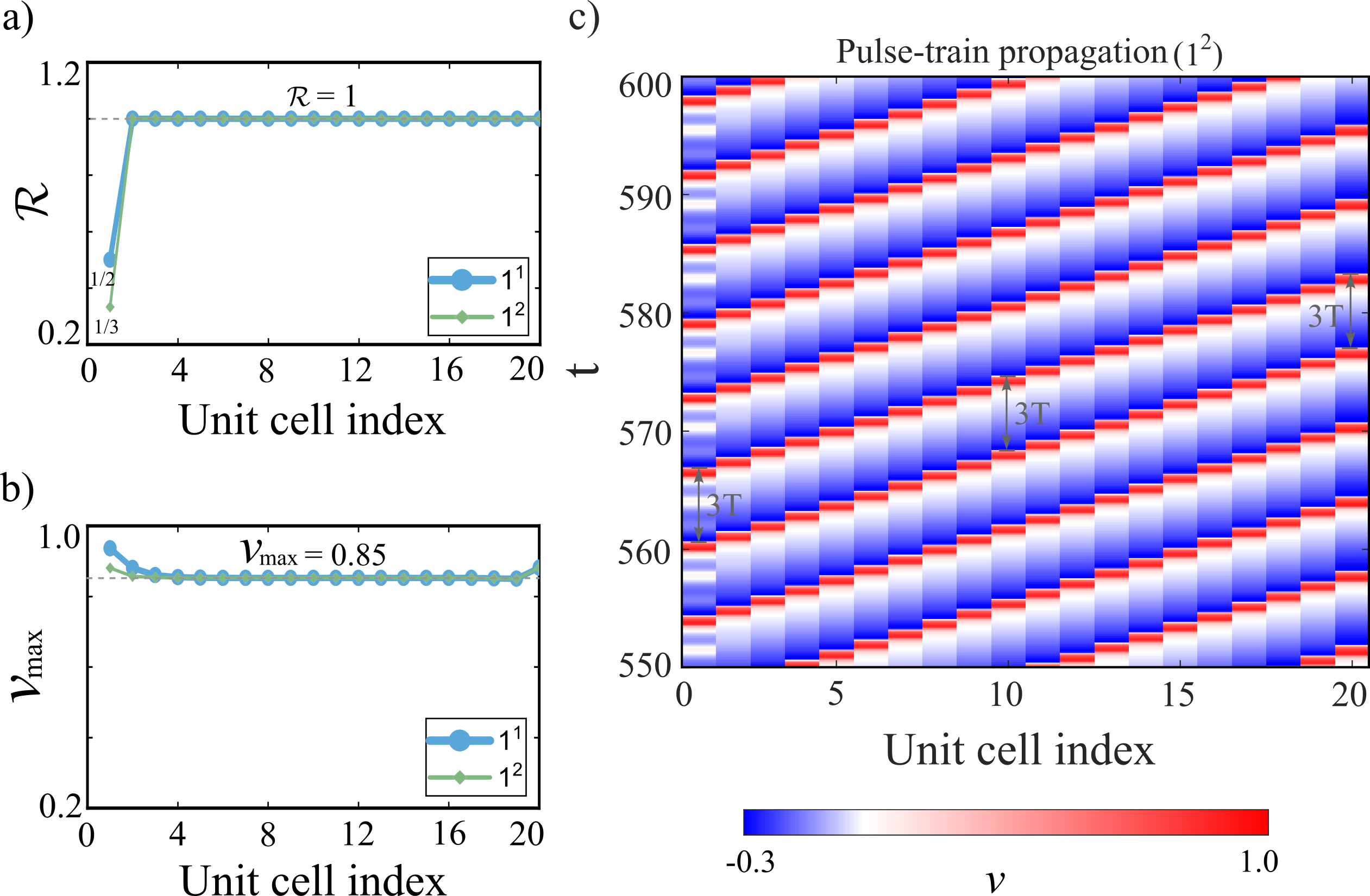}
	\caption{\label{Fig8} Propagation characteristics and spatial evolution of pulse trains in a one-dimensional lattice of coupled excitable mechanical units for $1^{s}$-type MMO states generated at the first unit. (a) Spike fraction, $\mathcal{R}$, calculated at each excitable unit for the $1^1$ and $1^2$ MMO states, respectively. The rapid convergence of $\mathcal{R}$ to unity indicates the suppression of small-amplitude oscillations and the persistence of only large-amplitude spikes during propagation. (b) Maximum spike amplitude, $v_{\mathrm{max}}$, at each excitable unit, demonstrating the rapid convergence toward pulses with constant amplitude after a short initial transient. The slight deviation observed at the last unit arises from finite-size boundary effects. (c) Spatiotemporal diagram illustrating the propagation of large-amplitude pulse trains with preserved temporal spacing along the lattice.}
\end{figure}

To quantify the spatial evolution of this behavior, we introduce the spike fraction, $\mathcal{R}$, defined as the ratio of the number of large-amplitude spikes to the total number of peaks within one response period at each excitable unit. This quantity, evaluated along the lattice and shown in Fig.~\ref{Fig8}(a), serves as a measure of the relative contribution of spike-like events compared to small-amplitude oscillations. The results presented in Fig.~\ref{Fig8}(a) correspond to the MMO states shown in Figs.~\ref{Fig7}(a) and \ref{Fig7}(b). At the first unit, $\mathcal{R}$ takes the values $1/2$ and $1/3$ for the $1^1$ and $1^2$ patterns, respectively, reflecting the coexistence of large-amplitude spikes and SAOs. However, from the second unit onward, $\mathcal{R}$ rapidly converges to unity, indicating that only large-amplitude spikes persist during propagation along the chain. This transition demonstrates the rapid emergence of a spike-dominated regime driven by the intrinsic filtering dynamics of the lattice. Moreover, the propagating pulse maintains an approximately constant amplitude and waveform during transmission along the chain. This behavior is evident from the temporal velocity profile of the last unit, where the large-amplitude spikes exhibit nearly identical amplitude and shape, indicating the consistency of the pulses reaching the end of the lattice. The spatial development of this stable propagation regime is further quantified in Figs.~\ref{Fig8}(b) and \ref{Fig8}(c), which present the maximum spike amplitude, $v_{\mathrm{max}}$, at each excitable unit and the corresponding spatiotemporal evolution, respectively. Figure~\ref{Fig8}(b) shows that, after a short initial transient, the spike amplitude rapidly converges to an approximately constant value along the lattice, indicating the propagation of pulses with constant amplitude. The slight deviation observed at the last unit arises from boundary effects associated with the finite size of the system.

In addition to amplitude preservation, the temporal spacing between successive large-amplitude spikes remains unchanged during propagation, as evidenced by the spatiotemporal diagram shown in Fig.~\ref{Fig8}(c). This invariance originates from the fact that the spikes are uniformly and sufficiently separated in time, allowing each downstream unit to fully recover from its refractory state before the arrival of the next excitation. Consequently, each incoming pulse consistently triggers a complete excitable response without noticeable distortion. Collectively, these results demonstrate that, for $1^{s}$-type MMOs, the excitable lattice supports robust and dispersion-free propagation of large-amplitude pulses while preserving the essential characteristics of the pulse train, including pulse amplitude and temporal spacing.

\subsection{$L^1$-type MMOs at the first unit}\label{subsecB}
\begin{figure}
	\centering
	\includegraphics[width=0.55\linewidth]{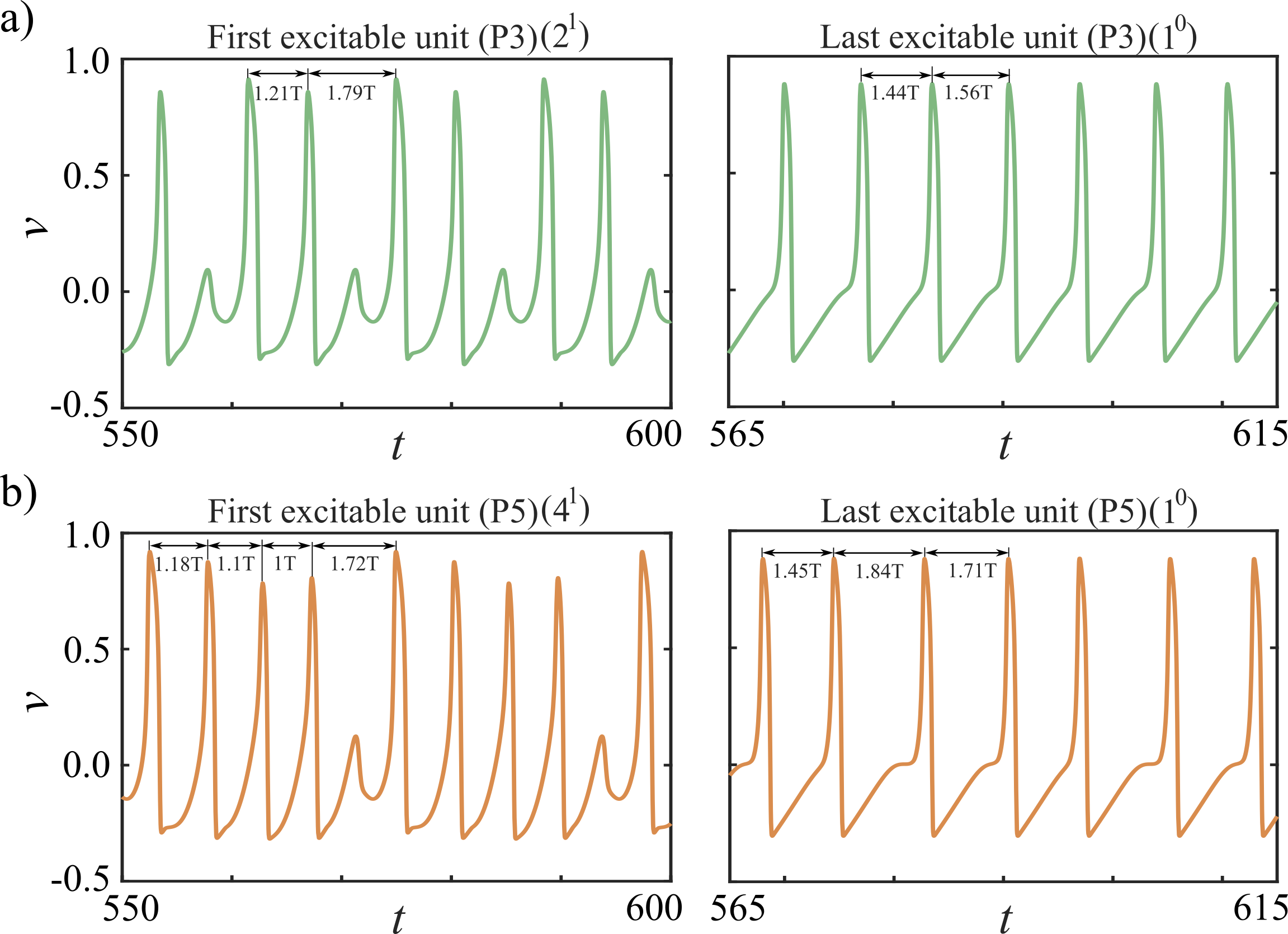}
	\caption{\label{Fig9} Temporal velocity profiles of the first and last units in a one-dimensional lattice of coupled excitable mechanical units ($N=20$) for $L^{1}$-type MMO states generated at the first unit. Panels (a) and (b) correspond to the forcing parameters $(\Omega,F)=(1.4,0.15)$ and $(1.4,0.17)$, representing the $2^{1}$ and $4^{1}$ patterns, respectively. The response at the first unit exhibits multiple large-amplitude spikes followed by a single small-amplitude oscillation within one response period. During propagation through the lattice, the small-amplitude oscillations are suppressed, and only large-amplitude spikes persist at the last unit. However, as shown in panel (b), not all large-amplitude spikes are successfully transmitted, indicating selective filtering induced by the refractory dynamics of the excitable medium.}
\end{figure}
\begin{figure}
	\centering
	\includegraphics[width=0.64\linewidth]{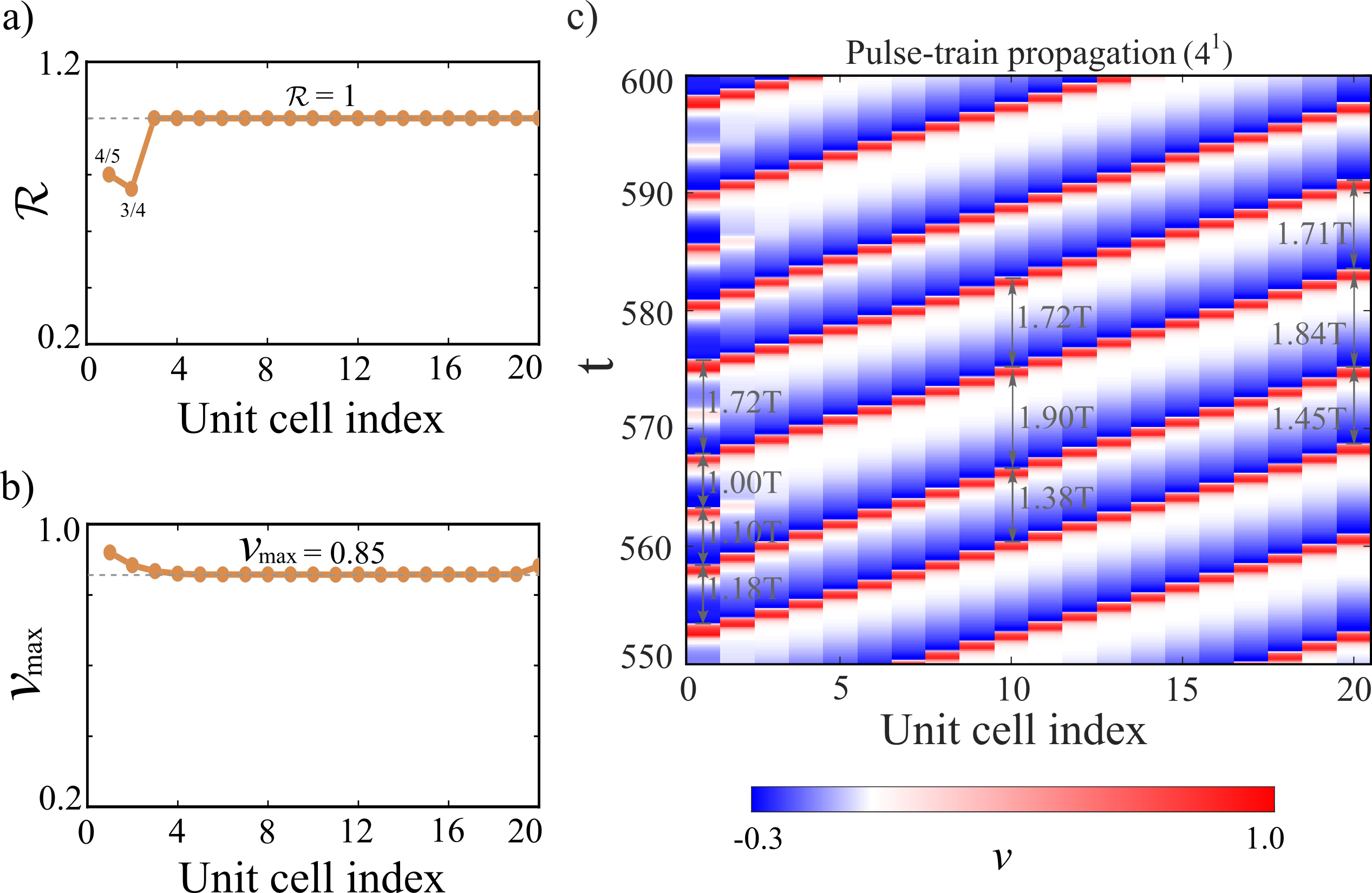}
	\caption{\label{Fig10} Spatial evolution and propagation characteristics of pulse trains in a one-dimensional lattice of coupled excitable mechanical units for $L^{1}$-type MMO states generated at the first unit. (a) Spike fraction $\mathcal{R}$ evaluated at each excitable unit for the $4^{1}$ MMO state shown in Fig.~\ref{Fig9}(b). (b) Maximum spike amplitude, $v_{\mathrm{max}}$, at each excitable unit, demonstrating rapid convergence toward a nearly constant value after a short initial transient. (c) Spatiotemporal diagram illustrating the propagation of large-amplitude pulses, highlighting selective spike transmission and the evolution of interspike spacing.}
\end{figure}
We next investigate pulse train propagation in the lattice when the first excitable unit exhibits $L^{1}$-type MMOs. Figures~\ref{Fig9}(a) and \ref{Fig9}(b) present the temporal velocity profiles of the first and last excitable units corresponding to the $2^{1}$ and $4^{1}$ patterns, respectively. In these states, the first unit exhibits multiple large-amplitude spikes followed by a single small-amplitude oscillation within one response period. Similar to the $1^{s}$-type MMOs discussed previously, the response at the last unit is dominated exclusively by large-amplitude spikes, while the SAOs are suppressed during propagation through the lattice. However, a key distinction emerges in that not all large-amplitude spikes generated at the first unit are successfully transmitted through the chain. As illustrated in Fig.~\ref{Fig9}(b), the third spike fails to propagate to the last unit. This suppression arises from the incomplete recovery of downstream excitable units from preceding excitation events, which prevents the incoming perturbation from triggering a full excitable response. Consequently, the closely spaced spike sequences associated with the $L^{1}$-type MMOs undergo selective filtering during propagation through the excitable medium. A more detailed characterization of this mechanism is provided in Fig.~\ref{Fig10}(a) for the $4^{1}$ MMO state. At the first unit, the ratio $\mathcal{R}=4/5$ reflects the presence of four large-amplitude spikes followed by one small-amplitude oscillation within a single response period. Upon propagation to the second unit, this ratio decreases to $\mathcal{R}=3/4$, as the small-amplitude oscillation is suppressed and the third spike is transformed into a subthreshold response due to the incomplete recovery of the receiving unit. From the third unit onward, the ratio rapidly converges to $\mathcal{R}=1$, indicating that only large-amplitude spikes persist while all small-amplitude components are eliminated. Consequently, the transmitted waveform asymptotically evolves into a spike-dominated sequence in the downstream lattice. This progressive evolution highlights a recovery-controlled filtering process, in which the local refractory state governs both the transmission and transformation of incoming excitations. We therefore identify this mechanism as \emph{refractory-mediated selective pulse transmission} in the lattice.
\begin{figure}
	\centering
	\includegraphics[width=0.56\linewidth]{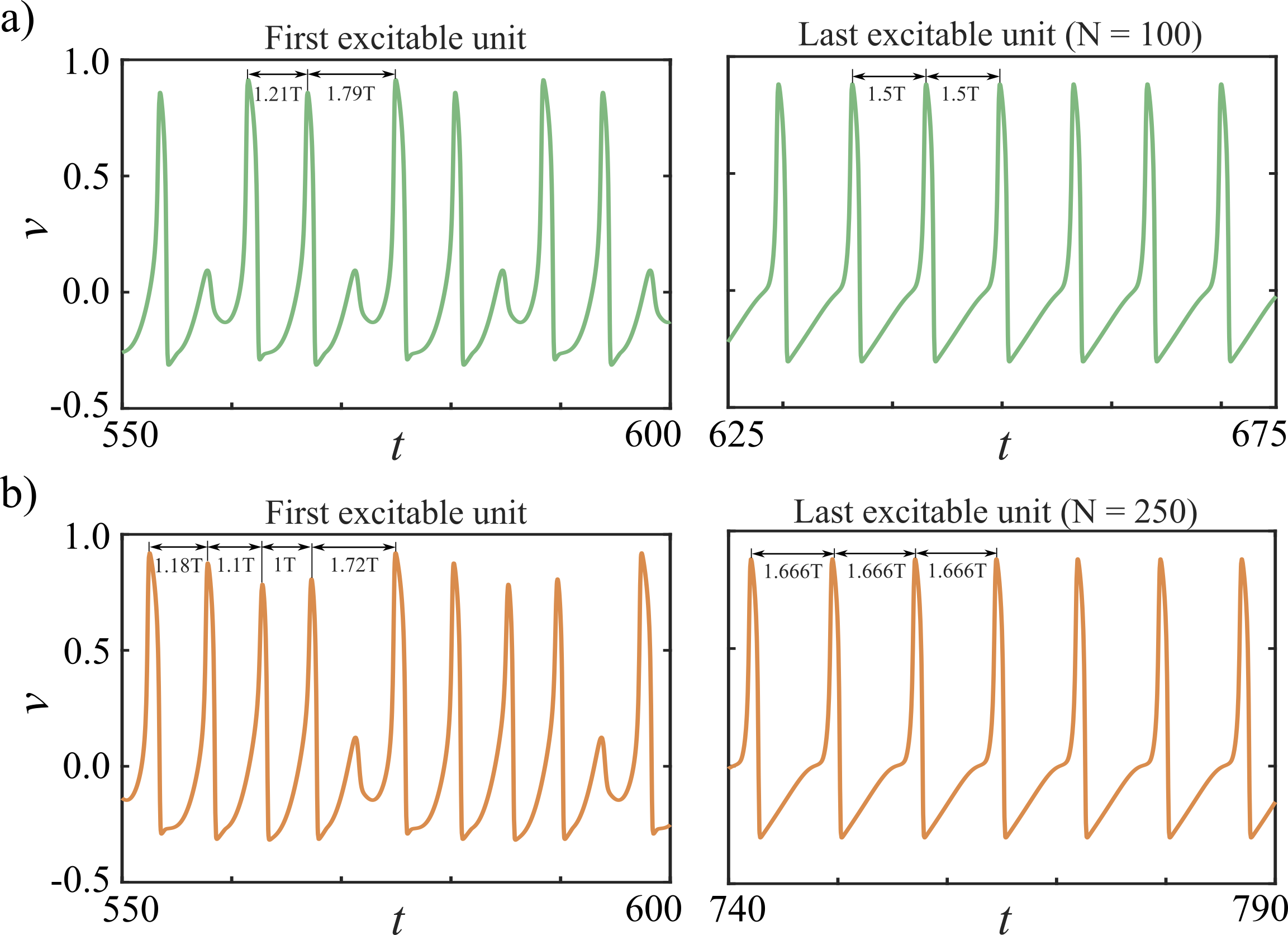}
	\caption{\label{Fig11} Regularization of interspike intervals during pulse train propagation in extended lattices of coupled excitable mechanical units, depicted through the temporal velocity profiles of the first and last units for (a) the $2^{1}$ MMO state with $N=100$ and (b) the $4^{1}$ MMO state with $N=250$. In both cases, the initially nonuniform interspike intervals progressively reorganize during propagation, leading to the emergence of pulse trains with uniform temporal spacing at the last unit.}
\end{figure} 

The large-amplitude pulses that remain after this filtering process maintain an approximately constant amplitude during propagation through the lattice, as evidenced by the temporal velocity profiles of the last unit shown in Figs.~\ref{Fig9}(a) and \ref{Fig9}(b). The spatial evolution of the pulse amplitude is quantified in Fig.~\ref{Fig10}(b), where the maximum spike amplitude rapidly converges to an approximately constant value after a short initial transient. However, in contrast to the $1^{s}$-type MMOs, the temporal spacing between successive large-amplitude spikes is not preserved during propagation, as observed in Figs.~\ref{Fig9}(a) and \ref{Fig9}(b), and further illustrated by the spatiotemporal diagram in Fig.~\ref{Fig10}(c). This distinction originates from the intrinsic temporal structure of the $L^{1}$ patterns, in which the large-amplitude spikes are nonuniformly distributed in time. Consequently, the recovery dynamics of the excitable medium between successive excitation events become uneven, causing downstream units to respond differently to incoming pulses depending on their instantaneous refractory state. Despite this initial temporal irregularity, the pulse train undergoes gradual reorganization during propagation through the lattice. Successive spikes experience unequal modifications to their effective propagation times, leading to a redistribution of interspike intervals along the chain. In particular, initially shorter intervals progressively increase, whereas initially longer intervals decrease, thereby reducing temporal disparity between consecutive spikes. For sufficiently long lattices, this self-adjusting process drives the system toward asymptotically periodic pulse propagation with uniform interspike intervals. This behavior is illustrated in Fig.~\ref{Fig11}(a)--(b), where simulations performed for extended lattices demonstrate the asymptotic emergence of uniform interspike intervals. Related phenomena have been reported in biological excitable media~\cite{villagran2013variations,bucher2011beyond,de2010alternans,elphick1990impulse}, particularly in single neurons of locusts and crayfish~\cite{villagran2013variations}, as well as in cultured strands of neonatal rat cardiac cells~\cite{de2010alternans}. Similar dynamics have also been observed in an optical excitable system with delayed feedback, where initially irregular pulse trains evolve toward stable periodic states~\cite{terrien2020equalization}. Although this system does not involve spatial propagation through a lattice, both frameworks exhibit a self-organized equalization of interspike intervals.

The dependence of this temporal reorganization on the lattice length $N$, further implies that the effective interspike interval can be controlled through the system size. Since the convergence toward uniform temporal spacing develops continuously during propagation, lattices with different values of $N$ exhibit different degrees of temporal regularization, thereby enabling tunability of the resulting pulse train timing. Consequently, the excitable mechanical lattice effectively functions as a geometry-dependent temporal filter, in which the output spike timing is governed by the intrinsic recovery dynamics and finite propagation distance. Collectively, these results demonstrate that, although $L^{1}$-type MMOs initially generate temporally nonuniform excitation patterns, the excitable mechanical lattice intrinsically reorganizes the propagating pulse train toward a stable and temporally ordered propagation regime, while simultaneously providing a mechanism for controllable spike-timing in extended excitable media.

\subsection{Composite MMO patterns and aperiodic states at the first unit}
\begin{figure}
	\centering
	\includegraphics[width=1\linewidth]{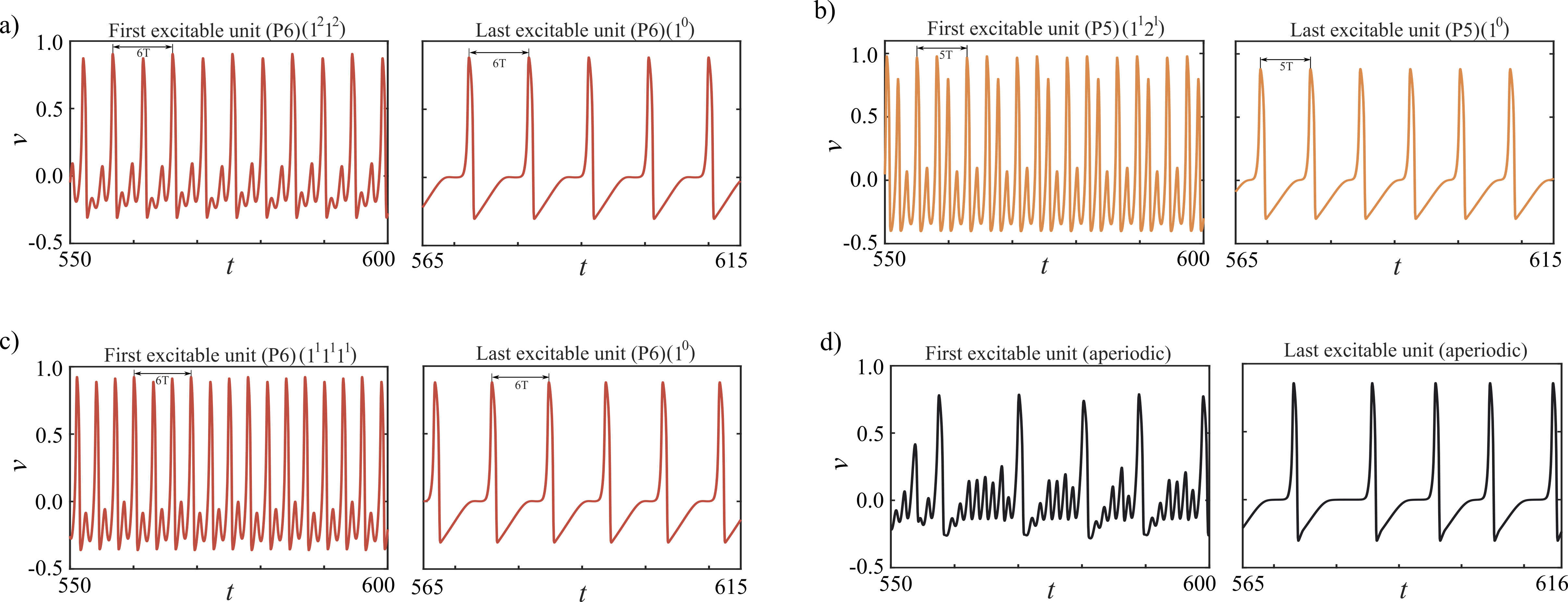}
	\caption{\label{Fig12} Temporal velocity profiles of the first and last units in a one-dimensional lattice of coupled excitable mechanical units ($N=20$) for composite MMO patterns and aperiodic states at the first unit. Panels (a)–(c) correspond to the composite MMO patterns $1^{2}1^{2}$, $1^{1}2^{1}$, and $1^{1}1^{1}1^{1}$, respectively, while panel (d) corresponds to aperiodic excitation at the first unit. The corresponding forcing parameters are (a) $(\Omega,F)=(4,0.2)$, (b) $(0.6,4)$, (c) $(4.2,0.4)$, and (d) $(5,0.12)$. In all cases, the dynamics at the last unit exhibit suppression of small-amplitude oscillations and selective transmission of large-amplitude spikes, leading to a simplified pulse train structure relative to the excitation at the first unit.}
\end{figure}

We now extend the analysis to pulse train propagation in the lattice when the first unit exhibits composite MMO patterns and aperiodic states. The composite MMO patterns are characterized by complex concatenations of distinct MMO motifs within a single response period. These states arise in parameter regimes associated with newly formed intricate structures in the response landscape, which emerge due to coupling-induced interactions within the excitable lattice.
Figures~\ref{Fig12}(a)--\ref{Fig12}(c) present the temporal velocity profiles of the first and last units corresponding to the composite MMO patterns $1^{2}1^{2}$, $1^{1}2^{1}$, and $1^{1}1^{1}1^{1}$, respectively, generated at the first unit. The temporal velocity profiles of the last unit demonstrate that the propagation dynamics retain the same qualitative features observed previously for the $1^{s}$- and $L^{1}$-type MMO states. In all three cases, the refractory-mediated selective pulse transmission mechanism persists, leading to the suppression of SAOs and the selective propagation of large-amplitude spikes through the lattice. Consequently, the pulse trains reaching the last unit exhibit a substantially simplified temporal organization compared to the more intricate excitation sequences generated at the first unit.

A qualitatively similar propagation behavior is observed when the first unit exhibits aperiodic dynamics, as shown in Fig.~\ref{Fig12}(d). In this regime, the excitation sequence at the first unit is characterized by irregular temporal spacing and aperiodic repeating of spike organization. Although the refractory-mediated filtering mechanism continues to suppress weak and closely spaced excitations during propagation, the absence of an underlying periodic structure prevents complete temporal regularization of the pulse train, even in sufficiently long lattices. As a result, the downstream dynamics retain signatures of temporal irregularity, demonstrating that the excitable lattice can partially reorganize but not fully regularize intrinsically aperiodic excitation patterns. 

\section{\label{Sec4} Conclusions\protect\\}
In this work, we first investigated a single excitable mechanical unit to characterize its dynamical regimes in the forcing-amplitude–frequency parameter space. The resulting response landscape revealed island-like domains of higher-period oscillations embedded within broader regions of period-one responses, alongside regions of aperiodic dynamics. Furthermore, the system exhibits a sharp transition between subthreshold and suprathreshold oscillatory responses upon a small change in the forcing amplitude, reflecting its intrinsic excitability. These contrasting dynamical regimes are separated by a transition boundary whose position varies nonmonotonically with the forcing frequency: the threshold amplitude initially decreases, reaches a minimum at intermediate frequencies, and subsequently increases. This nonmonotonic dependence highlights the interplay between the intrinsic dynamical timescales of the excitable unit and the timescale of the external forcing. A detailed analysis of the response across the forcing parameter space further revealed a rich variety of MMOs, characterized by alternating sequences of SAOs and large-amplitude spikes. These findings elucidate how periodic forcing organizes the dynamics of an individual excitable unit and provide insight into how the boundary unit of a spatially extended system may respond under periodic forcing.

We then investigated a one-dimensional excitable mechanical lattice subjected to periodic boundary forcing to show how the complex oscillatory dynamics of the first unit influence pulse propagation through the lattice. The first unit exhibits a rich variety of temporal patterns, including regular MMOs, characterized by periodic alternation of SAOs and large-amplitude spikes; composite MMOs, formed by concatenations of distinct MMO motifs; and irregular MMOs, characterized by nonuniform amplitudes and temporal spacing. Despite this complexity at the first unit, propagation through the lattice selectively suppresses SAOs and closely spaced spikes, giving rise to robust pulse trains composed of well-separated, large-amplitude spikes. This selective transmission is governed by the refractory dynamics of the lattice, which hinder the propagation of temporally clustered excitations. Furthermore, when the interspike intervals at the first unit are nonuniform, the lattice reorganizes the temporal structure of the transmitted pulse train. Specifically, the interspike intervals evolve systematically along the lattice as a consequence of the interplay between excitability and refractory dynamics in the constituent units. The extent of this temporal reorganization depends on the lattice length, providing a means of tuning the interspike intervals through system size. Consequently, lattices of different lengths exhibit different degrees of temporal regularization, with the spacing between successive spikes becoming more uniform during propagation. These results establish the excitable lattice as a nonlinear filter that transforms complex boundary-generated dynamics into more coherent propagating pulse trains and offers a mechanism for controlling spike timing in extended excitable media.

From a broader perspective, these findings may be relevant to other excitable media, including optical and electronic systems, in which similar mechanisms could provide a basis for controlling information-carrying pulse sequences and enabling wave-based computing functionalities. In particular, the observed dependence of the interspike interval on lattice length indicates a potential means of tuning the temporal spacing of successive pulses without directly manipulating the underlying excitation dynamics. Furthermore, refractory-mediated spike selection may suppress spurious excitations by preventing closely spaced pulses from propagating successfully, thereby providing a dynamical filtering mechanism that could enhance the robustness of signal transmission in the presence of noise or external perturbations. These findings motivate further investigations into the practical realization of such dynamics in mechanical systems and hybrid platforms, including optomechanical systems and other coupled physical architectures. Extending the present framework to these settings could enable the development of adaptive metamaterial-based wave filters for the controlled routing of energy and information, as well as neuromorphic information-processing schemes based on the propagation and interaction of pulse trains.

\printcredits

\section*{Declaration of competing interest}
The authors declare that they have no known competing financial interests or personal relationships that could have appeared to influence the work reported in this paper.

\section*{Acknowledgements}
The authors acknowledge financial support from the CNRS through its MITI interdisciplinary programs. K. P. gratefully acknowledges support from the postdoctoral fellowship funded by the Institut d’Acoustique-Graduate School (IA-GS).

\section*{Data availability}
All data that support the findings of this study are included within the article.

\appendix 
\section{Determination of coefficients in the dynamic friction function}\label{appA}
\begin{figure}
	\centering
	\includegraphics[width=0.5\linewidth]{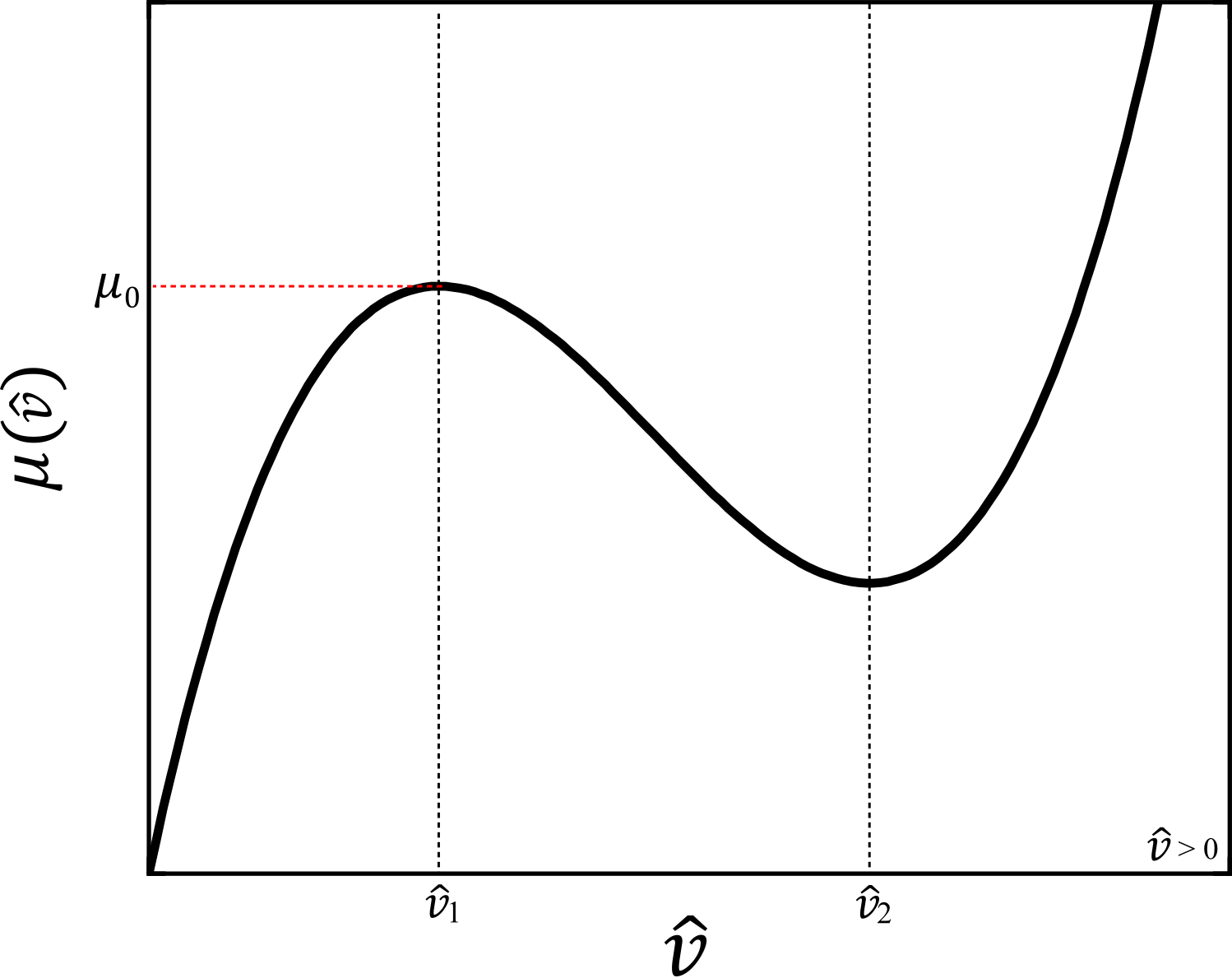}
	\caption{\label{FigA1}Schematic illustration of the cubic friction function $\mu(\hat{v})$ exhibiting an $N$-shaped dependence on the normalized relative velocity $\hat{v}$. The curve possesses a local maximum at $\hat{v}=\hat{v}_1$ and a local minimum at $\hat{v}=\hat{v}_2$, with $0<\hat{v}_1<\hat{v}_2$. The value of the local maximum is $\mu_0$.}
\end{figure}

In this Appendix, we derive the coefficients of the cubic friction function $\mu(\hat{v})$ such that it exhibits a smooth $N$-shaped dependence on the normalized relative velocity $\hat{v}$, as shown in Fig.~\ref{FigA1}. This behavior is characterized by the presence of a local maximum at $\hat{v}=\hat{v}_1$ and a local minimum at $\hat{v}=\hat{v}_2$, with $0<\hat{v}_1<\hat{v}_2$. These requirements are imposed by enforcing $\mu'(\hat{v}_1)=0$ and $\mu'(\hat{v}_2)=0$, where the prime denotes differentiation with respect to $\hat{v}$. For the cubic friction law $\mu(\hat{v}) = a |\hat{v}|^3 - b |\hat{v}|^2 + c |\hat{v}|$, it is sufficient to consider $\hat{v}>0$, for which the absolute value can be omitted without loss of generality. The derivative then takes the form
\begin{equation}
	\mu'(\hat{v}) = 3a \hat{v}^2 - 2b \hat{v} + c
	\label{EqA1}
\end{equation}
Imposing stationary points at $\hat{v}_1$ and $\hat{v}_2$ allows the derivative to be written in factored form as
\begin{equation}
	\mu'(\hat{v}) = 3a(\hat{v}-\hat{v}_1)(\hat{v}-\hat{v}_2)
	\label{EqA2}
\end{equation}
which, upon expansion and comparison of coefficients with the Eq.~\ref{EqA1}, yields
\begin{equation}
	b=\frac{3a}{2}(\hat{v}_1+\hat{v}_2), \qquad
	c=3a\,\hat{v}_1 \hat{v}_2
	\label{EqA3}
\end{equation}
The magnitude of the friction function at the local maximum is prescribed as $\mu(\hat{v}_1)=\mu_0>0$. Substituting the above expressions of $b$ and $c$ into $\mu(\hat{v})$ and evaluating at $\hat{v}=\hat{v}_1$ gives
\begin{equation}
	\mu(\hat{v}_1) = \frac{a \hat{v}_1^2}{2}\left(3\hat{v}_2 - \hat{v}_1\right)
	\label{EqA4}
\end{equation}
Equating this expression to $\mu_0$ determines the leading coefficient as
\begin{equation}
	a = \frac{2\mu_0}{\hat{v}_1^2\left(3\hat{v}_2 - \hat{v}_1\right)}
	\label{EqA5}
\end{equation}
The remaining coefficients then follow as
\begin{equation}
	b=\frac{3\mu_0(\hat{v}_1+\hat{v}_2)}{\hat{v}_1^2(3\hat{v}_2-\hat{v}_1)}, \qquad
	c=\frac{6\mu_0 \hat{v}_2}{\hat{v}_1(3\hat{v}_2-\hat{v}_1)}
	\label{EqA6}
\end{equation}
The obtained expressions for $a$, $b$, and $c$ guarantee that $\hat{v}_1$ and $\hat{v}_2$ correspond to a local maximum and a local minimum of $\mu(\hat{v})$, respectively, as confirmed by the sign of $\mu''(\hat{v})$ evaluated at these points for $0<\hat{v}_1<\hat{v}_2$. For illustration, we consider a representative set of parameters, $\mu_0 = 1$, $\hat{v}_1 = 0.4$, and $\hat{v}_2 = 1$, which satisfy $0<\hat{v}_1<\hat{v}_2$ and are sufficient to demonstrate the generic $N$-shaped structure of the friction function. Substituting these values into the expressions for the coefficients yields $a \approx 4.8$, $b \approx 10.1$, and $c \approx 5.8$, which are used throughout the manuscript.
\newpage
\section{Analogy with the FHN Model} \label{appB}
In this Appendix, we compare the proposed excitable mechanical unit with the FHN model in terms of their fast--slow structure and response characteristics. The governing equations of the excitable mechanical unit are written in first-order form in Eq.~\ref{Eq5}. Introducing the relative velocity $w = v - v_p$ and rescaling time according to $t = \tau/\gamma$, Eq.~\ref{Eq5} can be rewritten as
\begin{eqnarray}
	\frac{du}{dt} &=& \gamma (w + v_p),  \nonumber \\[2pt]
	\frac{dw}{dt} &=& -u - \Phi_{fr}(w)
	\label{EqB1}
\end{eqnarray}
For $\gamma\ll1$, Eq.~(\ref{EqB1}) has an explicit fast--slow structure, with $w$ evolving on the fast time scale and $u$ evolving on the slow time scale. To establish an analogy between the proposed excitable mechanical unit and the FHN model, we consider a commonly used form of the FHN model, given by
\begin{eqnarray}
	\frac{dy}{dt} &=& \epsilon (x + \kappa_1 - \kappa_2 y), \nonumber \\[2pt]
	\frac{dx}{dt} &=& -y - \Psi(x) + I
	\label{EqB2}
\end{eqnarray}
where $x$ and $y$ denote the fast activation and slow recovery variables, respectively. The parameter $\epsilon\ll1$ controls the separation between the fast and slow time scales, while $\kappa_1$ and $\kappa_2$ characterize the slow recovery dynamics. The function $\Psi(x)$ describes the nonlinear fast dynamics, and $I$ represents an external bias or input. For the standard FHN model, $\Psi(x)$ is typically chosen as a cubic function, producing the familiar folded, S-shaped critical manifold that provides the geometric framework for fast--slow excitable dynamics. 

A direct comparison of Eqs.~(\ref{EqB1}) and (\ref{EqB2}) shows that the correspondence is primarily geometrical. The mechanical relative velocity $w$ plays the role of the fast FHN variable $x$, while the displacement $u$ plays the role of the slow recovery variable $y$. The small parameter $\gamma$ plays a role analogous to $\epsilon$, establishing a separation between the fast evolution of $w$ and the slow evolution of $u$. The nonlinear friction force $\Phi_{fr}(w)$ determines the fast nullcline of the mechanical system in a manner analogous to the role of $\Psi(x)$ in the FHN model. In the mechanical system, the smooth, nonmonotonic $N$-shaped friction law is deliberately chosen such that the corresponding fast nullcline $u=-\Phi_{fr}(w)$ exhibits a folded, S-shaped geometry. In the singular limit $\gamma\rightarrow0$, this curve constitutes the critical manifold, along which the system undergoes slow evolution on the attracting branches, while rapid transitions occur when these branches lose normal hyperbolicity near the folds.

However, the slow-nullcline geometry differs fundamentally between the two systems. For the FHN model, the condition $dy/dt=0$ gives the inclined linear relation $y=(x+\kappa_1)/\kappa_2$, where $\kappa_2\neq0$, with $\kappa_1$ and $\kappa_2$ controlling the position and slope of the slow nullcline, respectively. In contrast, for the mechanical system, the condition $du/dt=0$ gives $w=-v_p$, yielding a vertical slow nullcline in the $(w,u)$ phase plane whose horizontal position is controlled by $v_p$, while its orientation remains fixed. This difference reflects the distinct forms of the slow dynamics in the two systems. In the FHN model, the slow evolution depends on both the fast and slow variables, whereas in the mechanical system it depends only on the fast variable and the constant $v_p$. Moreover, the roles of the bias parameters also differ between the two models. In the FHN system, the external bias $I$ shifts the fast nullcline and thereby modifies its intersection with the slow nullcline, providing a direct control of the fixed-point location. In the mechanical system, there is no separate additive parameter playing the same role as $I$; instead, the plate velocity $v_p$ shifts the slow nullcline horizontally and hence determines its intersection with the fast nullcline. Despite these differences, excitable dynamics in both systems depend on the position of the fixed point relative to the fast nullcline, thereby establishing a dynamical analogy between the excitable mechanical unit and the FHN model.

\section{Regime map of the excitable mechanical unit in the $(\Omega,F)$ parameter space}\label{appC}
\begin{figure}
	\centering
	\includegraphics[width=0.55\linewidth]{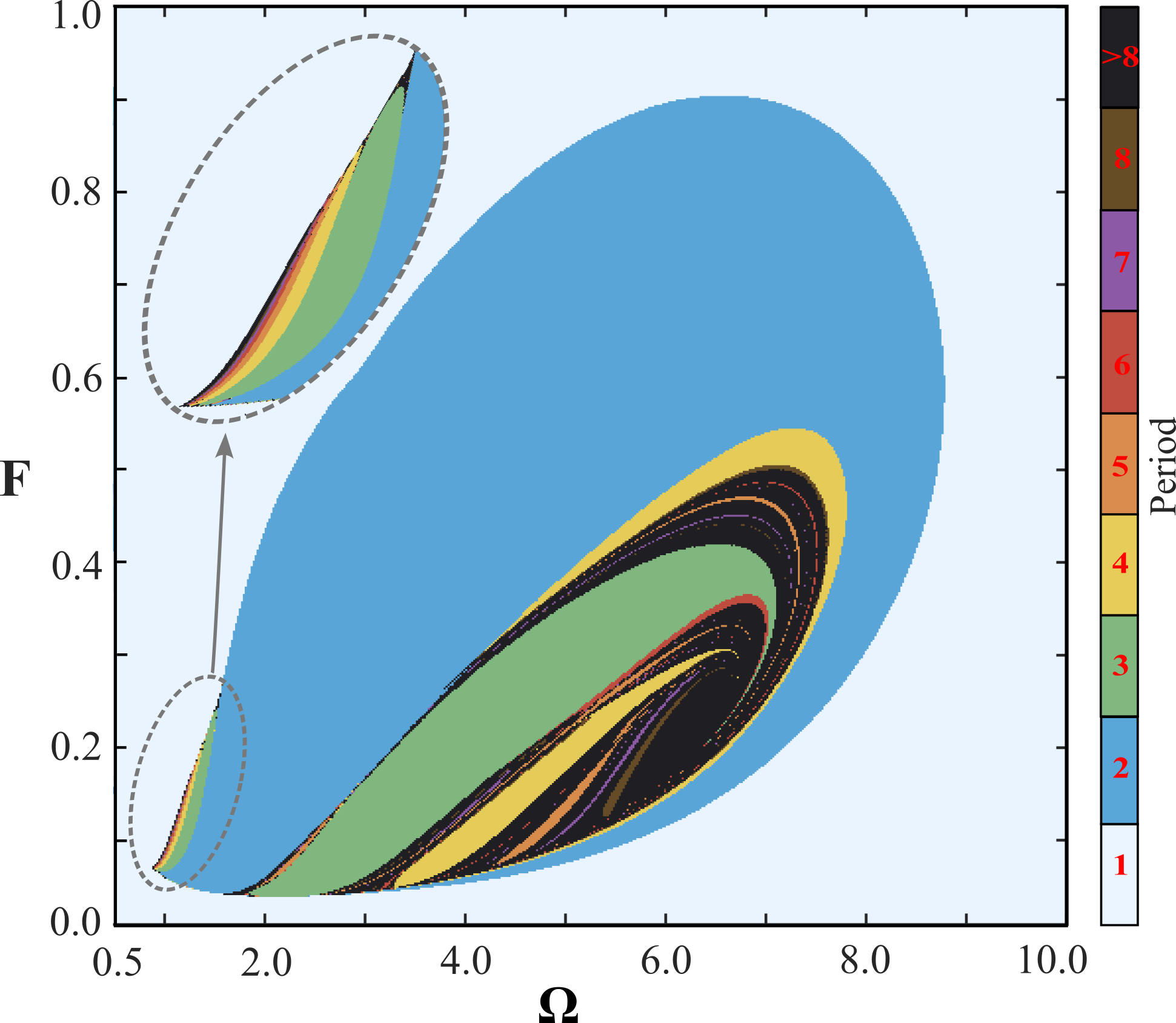}
	\caption{\label{FigA3} Regime map of the excitable mechanical unit in the $(\Omega,F)$ plane. Each point corresponds to the fundamental response period obtained from the stroboscopic Poincar\'e section after the removal of transients. The color scale represents the response period normalized by the forcing period, with integer periods explicitly labeled. The inset shows a magnified view of the region enclosed by the ellipse in the main panel.}
\end{figure}
In this Appendix, we present the regime map of the excitable mechanical unit in the $(\Omega,F)$ parameter space, providing a more detailed visualization of its periodic and aperiodic responses than that shown in Fig.~\ref{Fig3}. After discarding transients, the asymptotic response is analyzed using stroboscopic Poincar\'e sampling at the forcing period $T$. The number of distinct points in the resulting Poincar\'e section determines the fundamental response period. Repeating this procedure over the $(\Omega,F)$ parameter space yields the regime map shown in Fig.~\ref{FigA3}, where the color scale represents the response period normalized by the forcing period. Periodic responses from $P1$ to $P8$ are indicated by distinct colors, whereas black denotes higher-period responses ($P>8$) and aperiodic dynamics. As described in the main text, higher-period responses form an island-like domain embedded in the $P1$ background, with the $P2$ response forming the major portion of its boundary and the remaining higher-period responses interspersed with aperiodic dynamics. Most of the island-like domain is occupied by $1^s$-type MMOs. However, a small region in the lower-left portion of the domain, shown in the inset, is occupied by $L^1$-type MMOs.

\section{Response landscape of the first unit in a lattice of coupled excitable mechanical units}\label{appD}
\begin{figure}
	\centering
	\includegraphics[width=0.55\linewidth]{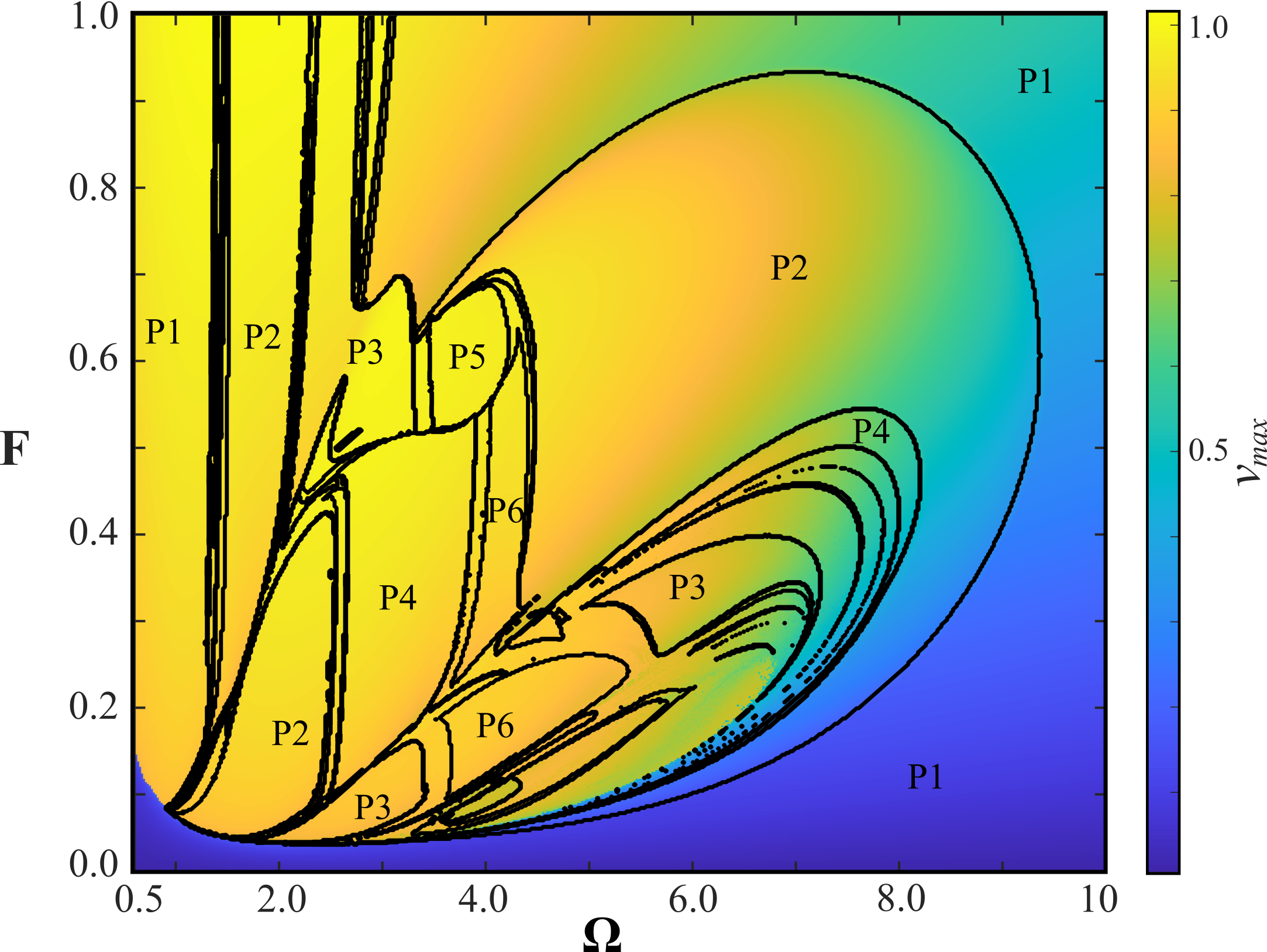}
	\caption{\label{FigA4} Response landscape of the first unit in a one-dimensional lattice of coupled excitable mechanical units ($N=20$, $\beta=0.2$) under sinusoidal forcing applied to the first unit, shown in the $(\Omega,F)$ parameter space. Compared with the single-unit case, the map exhibits a greater prevalence of higher-order subharmonic responses arising from interactions with neighboring units.}
\end{figure}
In this Appendix, we present the response landscape of the first unit in a one-dimensional lattice of $N$ coupled excitable mechanical units under periodic boundary forcing. Figure~\ref{FigA4} characterizes the local dynamical response of the first unit in the $(\Omega,F)$ parameter space and allows comparison with the corresponding response landscape of the periodically forced single excitable mechanical unit shown in Fig.~\ref{Fig3}. Compared with the isolated periodically forced unit, the coupled lattice exhibits a greater prevalence of higher-period responses and a more intricate arrangement of dynamical domains. This increased complexity arises from coupling-induced interactions between neighboring units. The local response of the first unit determines the temporal structure of the excitation injected into the lattice and therefore plays an important role in the subsequent propagation of pulse trains.

\bibliographystyle{cas-model1-names}
\bibliography{biblio}
\end{document}